\documentclass[prd,twocolumn,floatfix,preprintnumbers,showpacs]{revtex4}
\usepackage{graphicx}
\usepackage{dcolumn}
\usepackage{bm}
\usepackage{color}

\def\lsim{\mathrel{\mathop
  {\hbox{\lower0.5ex\hbox{$\sim$}\kern-0.8em\lower-0.7ex\hbox{$<$}}}}}
\def\gsim{\mathrel{\mathop
  {\hbox{\lower0.5ex\hbox{$\sim$}\kern-0.8em\lower-0.7ex\hbox{$>$}}}}}

\begin{document}

\newcommand{\1}{$\spadesuit$}
\newcommand{\half}{{1\over2}}
\newcommand{\nad}{n_{\rm ad}}
\newcommand{\niso}{n_{\rm iso}}
\newcommand{\ncor}{n_{\rm cor}}
\newcommand{\fiso}{f_{\rm iso}}
\newcommand{\ii}{\'{\'i}}
\newcommand{\bk}{{\bf k}}
\newcommand{\Ocdm}{\Omega_{\rm cdm}}
\newcommand{\ocdm}{\omega_{\rm cdm}}
\newcommand{\OM}{\Omega_{\rm M}}
\newcommand{\OB}{\Omega_{\rm B}}
\newcommand{\oB}{\omega_{\rm B}}
\newcommand{\OX}{\Omega_{\rm X}}
\newcommand{\cltt}{C_l^{\rm TT}}
\newcommand{\clte}{C_l^{\rm TE}}
\newcommand{\calR}{{\cal R}}
\newcommand{\calS}{{\cal S}}
\newcommand{\Rrad}{{\cal R}_{\rm rad}}
\newcommand{\Srad}{{\cal S}_{\rm rad}}
\newcommand{\calPR}{{\cal P}_{\cal R}}
\newcommand{\calPS}{{\cal P}_{\cal S}}

\input epsf

\preprint{IFT-UAM/CSIC-04-38, LAPTH-1065/04, astro-ph/0409326}
\title{Bounds on CDM and neutrino isocurvature perturbations from CMB 
and LSS data}
\author{Mar\'\i a Beltr\'an,$^1$ Juan Garc{\'\i}a-Bellido,$^1$ Julien 
Lesgourgues,$^2$ and Alain Riazuelo$^3$}
\affiliation{
$^1$Departamento de F\'\i sica Te\'orica \ C-XI, Universidad
Aut\'onoma de Madrid, Cantoblanco, 28049 Madrid, Spain\\
$^2$Laboratoire de Physique Th\'eorique LAPTH, F-74941
Annecy-le-Vieux Cedex, France\\
$^3$Institut d'Astrophysique de Paris, 98bis Boulevard Arago, 
75014 Paris, France
}
\date{September 14, 2004}
\pacs{98.80.Cq \\
Preprint \ IFT-UAM/CSIC-04-38, LAPTH-1065/04, astro-ph/0409326}
\begin{abstract}
Generic models for the origin of structure predict a spectrum of initial
fluctuations with a mixture of adiabatic and isocurvature
perturbations. Using the observed anisotropies of the cosmic microwave
backgound, the matter power spectra from large scale structure surveys
and the luminosity distance vs redshift relation from supernovae of type
Ia, we obtain strong bounds on the possible cold dark matter/baryon as
well as neutrino isocurvature contributions to the primordial
fluctations in the Universe. Neglecting the possible effects of spatial
curvature and tensor perturbations, we perform a Bayesian likelihood
analysis with thirteen free parameters, including independent spectral
indexes for each of the modes and for their cross-correlation angle.  We
find that around a pivot wavenumber of $k=0.05\,h\,$Mpc$^{-1}$ the
amplitude of the correlated isocurvature component cannot be larger than
about 60\% for the cold dark matter mode, 40\% for the neutrino density
mode, and 30\% for the neutrino velocity mode, at 2 sigma.
In the first case, our bound is larger than the WMAP first-year result,
presumably because we prefer not to include any data from Lyman-$\alpha$
forests, but then obtain large blue spectral indexes for the
non-adiabatic contributions.  We also translate our bounds in terms of
constraints on double inflation models with two uncoupled massive fields.
\end{abstract}

\maketitle

\section{Introduction}

With the increasing precision of the measurements of the cosmic
microwave background (CMB) anisotropies and large scale structures
(LSS) of the universe as well as various other astronomical
observations, it is now possible to have a clear and consistent
picture of the history and content of the universe since
nucleosynthesis.  Although the now widely used term of ``Standard
Model of Cosmology'' might remain premature as our knowledge of the
cosmological scenario is by far less precise than that of the Standard
Model of particle physics, the matter content of the universe as well
as its expansion rate are now known within a precision of a few
percents with great confidence. It is also well established that the
cosmological perturbations which gave rise to the CMB anisotropies and
the LSS of the universe was inflationary like, with a close to scale
invariant Harrison-Zeldovitch spectrum.  Moreover, the measurement of
both the temperature and polarization anisotropies of the cosmic
microwave background allows to test the paradigm of adiabaticity
of the cosmological perturbations and hence the precise nature of the
mechanism whih has generated them.

The simplest realizations of the inflationary paradigm predict an
approximately scale invariant spectrum of adiabatic and Gaussian
curvature fluctuations, whose amplitude remains constant outside the
horizon, and therefore allows cosmologists to probe the physics of
inflation through observations of the CMB anisotropies and the LSS
matter distribution. However, this is certainly not the only
possibility.  Models of inflation with more than one field generically
predict that, together with this so-called adiabatic component, there
should also be entropy, or isocurvature
perturbations~\cite{Linde:1985yf,Polarski:1994rz,Garcia-Bellido:1995qq,Gordon:2000hv,Wands:2002bn,Finelli:2000ya},
associated with fluctuations in number density between different
components of the plasma before photon decoupling, with a possible
statistical correlation between the adiabatic and isocurvature
modes~\cite{Langlois:1999dw}.  Baryon and cold dark matter (CDM)
isocurvature perturbations were proposed long ago~\cite{Efstathiou:1986}
as an alternative to adiabatic perturbations.  A few years ago, two
other modes, neutrino isocurvature density and velocity perturbations,
have been added to the list~\cite{Bucher:1999re}. 
Moreover, it is well known that entropy perturbations seed
curvature perturbations outside the
horizon~\cite{Polarski:1994rz,Garcia-Bellido:1995qq,Gordon:2000hv}, so
that it is possible that a significant component of the observed
adiabatic mode could be strongly correlated with an isocurvature
mode. Such models are generically called {\em curvaton
models}~\cite{Lyth:2001nq,Moroi:2001ct}, and are now widely studied as
an alternative to the standard paradigm. Furthermore, isocurvature modes
typically induce non-Gaussian signatures in the spectrum of primordial
perturbations.

In this paper, we describe in more detail the analysis performed in
Ref~\cite{Crotty:2003aa}, constraining the various isocurvature
components. We also extend it by including additional observational
constraints, and extra free parameters in the model. We use data from
the temperature power spectrum and temperature-polarization
cross-correlation measured by the WMAP satellite~\cite{WMAP}; as well
as from the small-scale temperature anisotropy probed by
VSA~\cite{VSA}, CBI~\cite{CBI} and ACBAR~\cite{ACBAR}; from the matter
power spectrum measured by the 2-degree-Field Galaxy Redshift Survey
(2dFGRS)~\cite{2dFGRS} and the Sloan Digital Sky Survey
(SDSS)~\cite{SDSS}; and also use data from the recent type Ia
Supernova compilation of Ref.~\cite{Riess2004}.  We do not use the
data from Lyman-$\alpha$ forests, since they are based on non-linear
simulations carried under the assumption of adiabaticity.  The first
bounds on isocurvature perturbations assumed uncorrelated
modes~\cite{Stompor:1995py}, but recently also correlated ones were
considered in
Refs.~\cite{Trotta:2001yw,Amendola:2001ni,Gordon:2002gv,Valiviita:2003ty,
Crotty:2003aa, Bucher:1999re, Moodley:2004ws}. Our general analysis
includes this possibility.  In the first part of this work, we will
not assume any specific model of inflation, nor any particular
mechanism to generate the perturbations (late decays, phase
transitions, cosmic defects, etc.), and thus will allow all five modes
--- adiabatic (AD), baryon isocurvature (BI), CDM isocurvature (CDI),
neutrino isocurvature density (NID) and neutrino isocurvature velocity
(NIV) --- to be correlated (or not) with each other , and to have
arbitrary tilts.  

In terms of model building, the simplest situation beyond the paradigm
of adiabaticity is that of a single isocurvature mode mixed with the
adiabatic one. Therefore, we shall not consider more than one
isocurvature mode at a time, and our primordial perturbations will be
described by three amplitudes and three spectral indices, associated
respectively with the adiabatic, isocurvature and cross-correlated
components. This choice is somewhat different from that of
Refs.~\cite{Trotta:2001yw,Bucher:1999re,Moodley:2004ws}, who introduce
several modes at a time, but a single tilt for each power spectrum of
primordial perturbations. The assumption that all the modes have
comparable amplitudes and a common tilt are both difficult to motivate
theoretically and, to our knowledge, all proposed mechanisms based on
inflation stand far from this case. For instance, double inflation leads
to at most one isocurvature mode, always with a tilt differing from the
adiabatic one; on the other hand, the curvaton scenario predicts a
single tilt, but only one isocurvature mode, fully correlated or
anti-correlated with the adiabatic mode.  However, in the absence of any
theoretical prior, we believe the approach of
Refs.~\cite{Trotta:2001yw,Bucher:1999re,Moodley:2004ws} is interesting
and complementary to ours.

For simplicity, we will neglect the possible effects of spatial
curvature and tensor perturbations, and assume that neutrinos are
massless~\cite{Pastor}.  Each model will be described by 
eleven cosmological parameters: the six usual parameters of the
standard $\Lambda$CDM model; the amplitude and spectral index of the
primordial isocurvature perturbation; the amplitude and spectral index
of the cross-correlation angle between the adiabatic and isocurvature
modes; and finally, a parameter $w$ describing the equation of state
of dark energy, assumed to be time-independent as a first
approximation. In addition, we will treat conservatively
the matter-to-light bias of the 2dF and SDSS redshift surveys as two
extra free parameters.  

In section II we describe the notation we use in our analysis of
isocurvature modes and discuss the relation with multi-field inflation,
and in particular two-field models, on which we will concentrate
ourselves. In section III we discuss the general bounds on our full
parameter space from CMB, LSS and SN data using a Bayesian likelihood
analysis. In section IV we analyze those bounds in a concrete model of
two-field inflation: double inflation, with two uncoupled massive
fields. A particular case is that in which the two fields have equal
masses, like in complex field inflation, which we show is not
ruled out. In section V we present our general conclusions.

\section{Notations}

For the theoretical analysis, we will use the notation and some of the
approximations of Refs.~\cite{Gordon:2000hv,Wands:2002bn}. During
inflation more than one scalar field could evolve sufficiently slowly
that their quantum fluctuations perturb the metric on scales larger
than the Hubble scale during inflation. These perturbations will later
give rise to one adiabatic mode and several isocurvature modes. We
will restrict ourselves here to the situation where there are only two
fields, $\phi_1$ and $\phi_2$, and thus only one isocurvature and one
adiabatic mode. Introducing more fields would complicate the
inflationary model and even then, it would be rather unlikely that
more than one isocurvature mode contributes to the observed
cosmological perturbations.  

The evolution during inflation will draw a trajectory in field space.
Perturbations along the trajectory (i.e. in the number of $e$-folds $N$)
will give rise to curvature perturbations on comoving hypersurfaces,
\begin{equation}
\calR_k = \delta N_k = H\,\delta t_k = H\,{\delta\rho_k\over
\dot\rho}\,,
\end{equation} 
while perturbations orthogonal to the trajectory will give rise to gauge
invariant entropy (isocurvature) perturbations,
\begin{equation}
\calS = \delta\ln{n_i\over n_j} = {\delta\rho_i\over(\rho_i+p_i)} -
{\delta\rho_j\over(\rho_j+p_j)}\,.
\end{equation} 
For instance, entropy perturbations in
cold dark matter during the radiation era can be computed as
$\calS_{\rm CDI} = \delta_{\rm cdm} - 3\delta_{\gamma}/4$.
In order to relate these perturbations during the radiation era with 
those produced during inflation, one has to follow the evolution
across reheating. 

During inflation we can always perform an
instantaneous rotation along the field trajectory and relate
the gauge-independent perturbations in the fields~\cite{Bardeen}, 
$\delta\hat{\phi}_i = \delta\phi_i + (\dot{\phi}_i/H) \psi$, with
perturbations along and orthogonal to the trajectory,
$\delta\sigma$ and $\delta s$,
\begin{equation}
\left(\begin{array}{c}\delta\sigma\\[2mm] \delta s\end{array}\right) =
\left(\begin{array}{rr}\cos\varphi & \sin\varphi\\[2mm] -\sin\varphi & 
\cos\varphi\end{array}\right)
\left(\begin{array}{c}\delta\hat{\phi}_1\\[2mm] 
\delta\hat{\phi}_2\end{array}\right)\,,
\end{equation}
with $\varphi$ the rotation angle between the two frames.

In this case, the curvature and entropy perturbations on
super-horizon scales can be written, in the slow-roll approximation,
as~\cite{Garcia-Bellido:1995qq,Gordon:2000hv}
\begin{eqnarray}
&&\calR_k = H\,{\dot\phi_1\delta\hat{\phi}_1+\dot\phi_2\delta\hat{\phi}_2\over
\dot\phi_1^2+\dot\phi_2^2} = H\,{\delta\sigma_k\over\dot\sigma}\,, \\[2mm]
&&\,\calS_k = \frac{2}{3}\dot{\varphi}\,{\dot\phi_1\delta\hat{\phi}_2-
\dot\phi_2\delta\hat{\phi}_1\over\dot\phi_1^2+\dot\phi_2^2} = 
\frac{2}{3}\dot{\varphi}\,{\delta s_k\over\dot\sigma} \,. 
\end{eqnarray}
where $\dot{\sigma}^2 \equiv \dot\phi_1^2+\dot\phi_2^2$.
The problem however is that, contrary to the case of purely adiabatic
perturbations, the amplitudes of both curvature and entropy perturbations
do not remain constant on superhorizon scales, but evolve with
time~\cite{Polarski:1994rz,Garcia-Bellido:1995qq}. In particular, due to
the conservation of the energy momentum tensor, the entropy perturbations
seed the curvature perturbations, and thus their amplitude during the
radiation dominated era evolves according 
to~\cite{Garcia-Bellido:1995qq,Gordon:2000hv}
\begin{eqnarray}\label{evolution}
&&\dot\calR_k = \alpha(t)H\,\calS_k\,, \\[2mm]
&&\,\dot\calS_k = \beta(t)H\,\calS_k\,,
\end{eqnarray}
where $\alpha$ and $\beta$ are time-dependent functions characterizing 
the evolution during inflation and radiation eras. A formal solution
can be found in terms of a transfer matrix, relating the amplitude at
horizon crossing during inflation (*) with that at a later time during
radiation,
\begin{equation}
\left(\begin{array}{c} \calR \\[2mm] \calS \end{array}\right) =
\left(\begin{array}{lr} 1 & T_{\calR\calS}\\[2mm]
 0 & T_{\calS\calS}\end{array}\right)
\left(\begin{array}{c} \calR_* \\[2mm] \calS_* \end{array}\right)\,,
\end{equation}
where the transfer functions are given by~\cite{Gordon:2000hv}
\begin{eqnarray}\label{TF}
&&T_{\calS\calS}(t,t_*) = 
\exp\left[\int_{t_*}^t\beta(t')H(t')dt'\right]\,, \\[2mm]
&&T_{\calR\calS}(t,t_*) = 
\int_{t_*}^t\alpha(t')H(t')T_{\calS\calS}(t',t_*)dt'\,.
\end{eqnarray}
Note that in the absence of primordial isocurvature perturbation,
$\calS_*=0$, the curvature perturbation remains constant and no
isocurvature perturbation is generated during the evolution. This is
the reason for the entries $T_{\calR\calR}=1$ and $T_{\calS\calR}=0$,
respectively, in the transfer matrix. Note also that in many models,
$\alpha(t)$ and $\beta(t)$ vanish after reheating, so that $(\calR_k,
\calS_k)$ remain constant during radiation domination on super-horizon
scales.  However, this is not true for instance when the fluid
carrying the isocurvature perturbations has a significant background
density (compared to the total Universe density), as assumed in the
curvaton scenario.

Since $\phi_1$ and $\phi_2$ are essentially massless during inflation,
we can treat them as free fields whose fluctuations at horizon
crossing have an amplitude $\delta\hat{\phi}_i =
(H_k/\sqrt{2k^3})\,e_i(\bk)$, where $H_k$ is the rate of expansion at
the time the perturbation crossed the horizon $(k_*=aH)$, and
$e_i(\bk)$ are Gaussian random fields with zero mean, $\langle
e_i(\bk)\rangle=0$ and $\langle e_i(\bk)e_j^*(\bk')\rangle
=\delta_{ij}\,\delta(\bk-\bk')$. Now, since $\delta\sigma_k$ and
$\delta s_k$ and just rotations of the field fluctuations, they are
also Gaussian random fields of amplitude $H_k/\sqrt{2k^3}$. However,
the time evolution (\ref{evolution}) will mix those modes and will
generically induce correlations and non Gaussianities.

Therefore, the two-point correlation function or power spectra of both
adiabatic and isocurvature perturbations, as well as their
cross-correlation, can be parametrized with three power laws, i.e.
three amplitudes and three spectral indices,
\begin{eqnarray}\nonumber
\Delta_\calR^2(k)\!&\equiv&\!
{k^3\over2\pi^2}\langle\calR^2\rangle = 
A^2\,\left({k\over k_0}\right)^{\nad-1}\,,\\[2mm]
\Delta_\calS^2(k)\!&\equiv&\!
{k^3\over2\pi^2}\langle\calS^2\rangle = 
B^2\,\left({k\over k_0}\right)^{\niso-1}\,,\\[2mm]
\Delta_{\calR\calS}^2(k)\!&\equiv&\!{k^3\over2\pi^2}\nonumber
\langle\calR\calS\rangle \, \nonumber\\[2mm]
&=& A\,B\,\cos\Delta_{k_0} 
\left({k\over k_0}\right)^{\ncor+\half(\nad+\niso)-1}\,,\nonumber
\end{eqnarray}
where $k_0$ is some pivot scale. 
Since the time
of horizon crossing 
$t_*$ in the transfer functions (\ref{TF}) is scale-dependent,
the correlation angle $\Delta(k)$ is in general a function of $k$.
In the above definitions, we
approximated  $\cos\Delta(k)$ by a power law 
with amplitude $\cos\Delta_{k_0}$
and tilt $\ncor$. So, we assumed implicitly that the inequality
\begin{equation}
\label{ncorrconst}
\left|\cos\Delta_{k_0}\right| \left({k\over k_0}\right)^{\ncor} \leq 1
\end{equation}
holds over all relevant scales. In the following analysis, we will
impose that for each value of $\cos\Delta_{k_0}$ the tilt $\ncor$
is restricted to the interval in which the inequality holds
between $k_{\rm min} = 4 \times 10^{-5}$Mpc$^{-1}$ and
$k_{\rm max} = 0.5\,$Mpc$^{-1}$, which is roughly 
the range probed by our CMB and LSS data sets.
Far from that range, we expect that next-order terms become
important and that the power-law approximation breaks down.
The fact that $\ncor$ can be non-zero was already considered in
a recent paper~\cite{Valiviita:2003ty}.


The angular power spectrum of temperature and polarization anisotropies
seen in the CMB today can be obtained from the radiation transfer
functions for adiabatic and isocurvature perturbations, $\Theta_l^{\rm
ad}(k)$ and $\Theta_l^{\rm iso}(k)$, computed from the initial conditions
$(\Rrad,\Srad) =(1,0)$ and $(0,1)$, respectively, and convolved with the
initial power spectra,
\begin{eqnarray}\nonumber
C_l^{\rm ad}\!&\equiv&\,\int {dk\over k}\,
\left[\Theta_l^{\rm ad}(k)\right]^2\,
\left({k\over k_0}\right)^{\nad-1}\,,\\[2mm]
C_l^{\rm iso}\!&\equiv&\,\int {dk\over k}\,
\left[\Theta_l^{\rm iso}(k)\right]^2\,
\left({k\over k_0}\right)^{\niso-1}\,,\nonumber\\[2mm]
C_l^{\rm cor}\!&\equiv&\,\int {dk\over k}\,
\Theta_l^{\rm ad}(k)\,\Theta_l^{\rm iso}(k)\,
\left({k\over k_0}\right)^{\ncor+\half(\nad+\niso)-1}\,,\nonumber
\end{eqnarray}
Then, the total angular power spectrum reads
\begin{equation}
C_l = A^2 \, C_l^{\rm ad} + B^2\,C_l^{\rm iso} + 
2\,A\,B\,\cos\Delta_{k_0}\,C_l^{\rm cor}\,.
\end{equation}
In many works (see for instance~\cite{Amendola:2001ni,WMAP}),
the following parametrization is employed:
\begin{equation}
C_l = C_l^{\rm ad} + {\cal B}^2\,C_l^{\rm iso} + 
2{\cal B}\,\cos\Delta_{k_0}\,C_l^{\rm cor}\,,
\end{equation}
where the global normalization has been marginalized over, and 
${\cal B}$ is the entropy to curvature perturbation ratio during 
the radiation era, ${\cal B}\equiv\Srad/\Rrad$. We will use here 
a slightly different notation,
used before by other groups~\cite{Langlois:1999dw,Stompor:1995py},
where $A^2 \equiv (1-\alpha)$ and $B^2\equiv\alpha$ (up to a global
normalization factor), so that
\begin{equation}\label{alphanotation}
C_l = (1-\alpha)\,C_l^{\rm ad} + \alpha\,C_l^{\rm iso} + 2\beta\,
\sqrt{\alpha(1-\alpha)}\,C_l^{\rm cor}\,.
\end{equation}
The parameter $\alpha$ runs from purely adiabatic ($\alpha=0$) to
purely isocurvature ($\alpha=1$), while $\beta$ defines the
correlation coefficient, with $\beta=+1(-1)$ corresponding to maximally
correlated(anticorrelated) modes. There is an obvious relation between
both parametrizations:
\begin{equation}
\alpha = {\cal B}^2/(1+{\cal B}^2)\,, \hspace{1cm} \beta = \cos\Delta_{k_0}\,.
\end{equation}
This notation has the advantage that the full parameter space of
$(\alpha,\ 2\beta\sqrt{\alpha(1-\alpha)})$ is contained within a
circle of radius $1/2$. The North and South rims correspond to fully
correlated ($\beta=+1$) and fully anticorrelated ($\beta=-1$)
perturbations, with the equator corresponding to uncorrelated
perturbations ($\beta=0$).  The East and West correspond to purely
isocurvature and purely adiabatic perturbations, respectively. Any
other point within the circle is an arbitrary admixture of adiabatic
and isocurvature modes.

 Note that interpreting the results in non fully correlated and anti
correlated models is made difficult by the fact that as soon as $\ncor
\neq 0$, the contours one obtains depend on the choice of the pivot
scale. For example in the simple case where $\nad = \niso$, the
relative amplitude between $C_l^{\rm ad}$ and $C_l^{\rm iso}$ is
unchanged when one changes the pivot scale, and hence $\alpha$ is
unchanged, but the correlation angle $\Delta (k)$, and hence $\beta$
changes with the pivot scale as $\beta(k_1) = \beta(k_0) (k_1 /
k_0)^{\ncor}$. In our example, if $\ncor > 0$, points within the
$(\alpha,\ 2\beta\sqrt{\alpha(1-\alpha)})$ circle are shifted
vertically toward the edge of the circle when one increases the pivot
scale and shifted toward the horizontal $\beta = 0$ line when one
decreases the pivot scale. 

We should emphasize that the three amplitude parameters $\ln[10^{10}
{\cal R}_{\rm rad}]$, $\alpha$ and $\beta$ are defined at $k = k_0$, and
that comparing bounds from various papers is straighforward only when
the pivot scale is the same. Throughout this paper, we will use $k_0 =
0.05$~Mpc$^{-1}$, which differs from our previous work
\cite{Crotty:2003aa}, but has the advantage of matching most of the
literature. This value of $k$ corresponds roughly to multipole number
$l_0 \sim 300$.

\section{General results}

We computed the Bayesian likelihood of each cosmological parameter
with a Monte Carlo Markov Chain method, using the public code
CosmoMC~\cite{cosmomc} with option MPI in the COSMOS supercomputer.
For each case, we ran 32 Markhov chains, and obtained between 30,000
and 60,000 samples (without including the multiplicity of each point).

We used the public code CAMB~\cite{camb} in order to compute the
theoretical prediction for the $C_l$ coefficients of the temperature and
polarization power spectra, as well as the matter spectra $P(k)$, for
all four different components (The BI mode is just a rescaling of the
CDI mode, see below). Moreover, we slightly modified the interface
between CAMB and CosmoMC in order to include the cross-correlated power
spectra, as well as three independent tilts $-$ in the CosmoMC jargon,
the three tilts and the three amplitudes were implemented into the code
as ``fast parameters'', in order to save a considerable amount of
time. The likelihood of each model was then computed using the detailed
information provided on the experimental websites (or directly from the
corresponding routines of the CosmoMC code when available), using three
groups of data sets: i) CMB data: 1398 points from WMAP ($2\leq
l\leq900$)~\cite{WMAPlike}, 8 points from VSA ($580\leq
l\leq1700$)~\cite{VSA}, 8 points from CBI ($700\leq
l\leq1800$)~\cite{CBI}, 7 points from ACBAR ($920\leq
l\leq1960$)~\cite{ACBAR}; ii) LSS data: 32 points from 2dFGRS (up to
$k_{\rm max} = 0.1\ h$ Mpc$^{-1}$)~\cite{2dFGRS}, 17 points from SDSS
(up to $k_{\rm max} = 0.15\ h$ Mpc$^{-1}$)~\cite{SDSS}; and iii) SNIa
data: 157 points from Riess et al.~\cite{Riess2004}. We have checked
that the inclusion of the HST~\cite{Freedman:2000cf} and BBN priors are
irrelevant for the determination of the Hubble parameter and the baryon
density; that is, the data from CMB, LSS and SNIa is enough in order to
determine these parameters, and therefore we ignored the priors in the
final analysis.

\begin{figure*}[ht]
\begin{center}
\includegraphics[angle=-90,width=8.5cm]{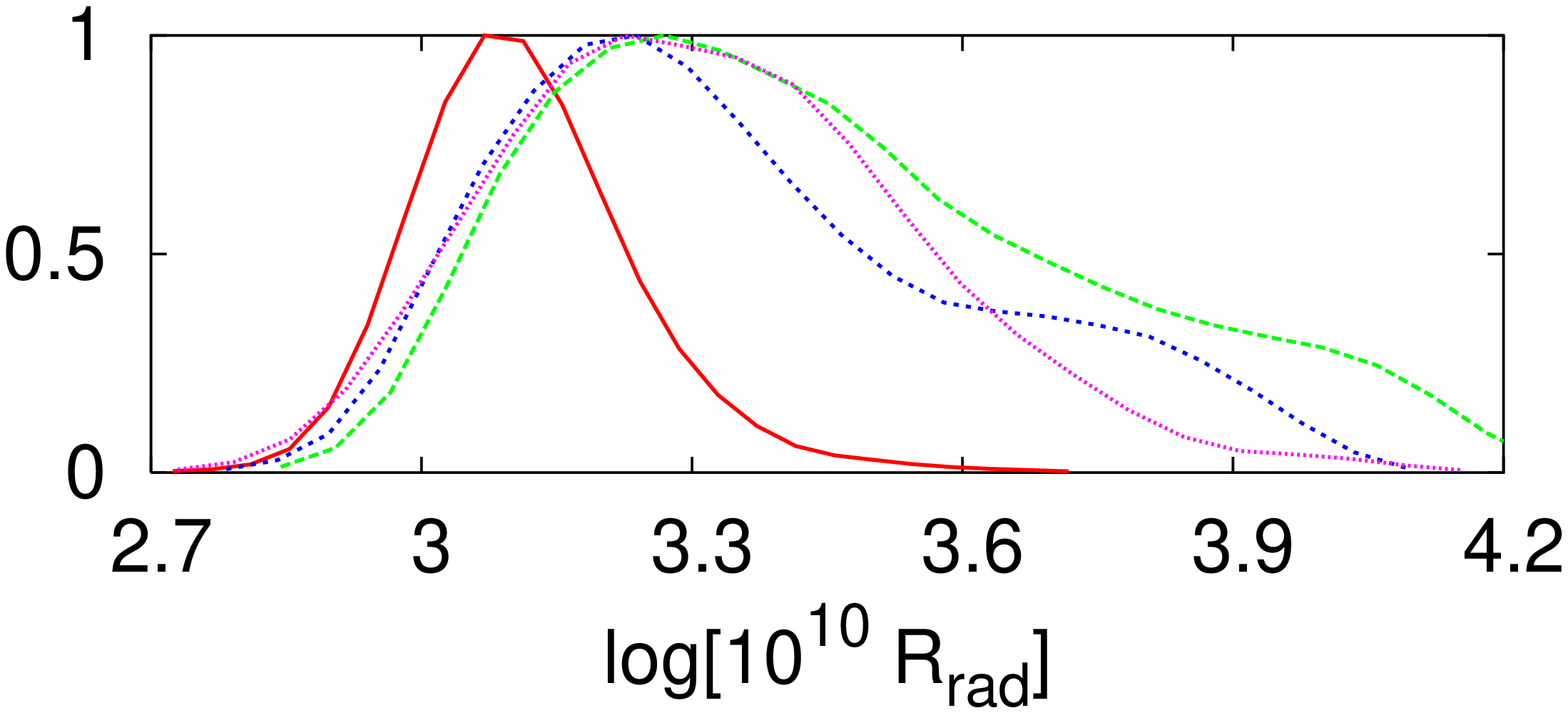}
\includegraphics[angle=-90,width=8.5cm]{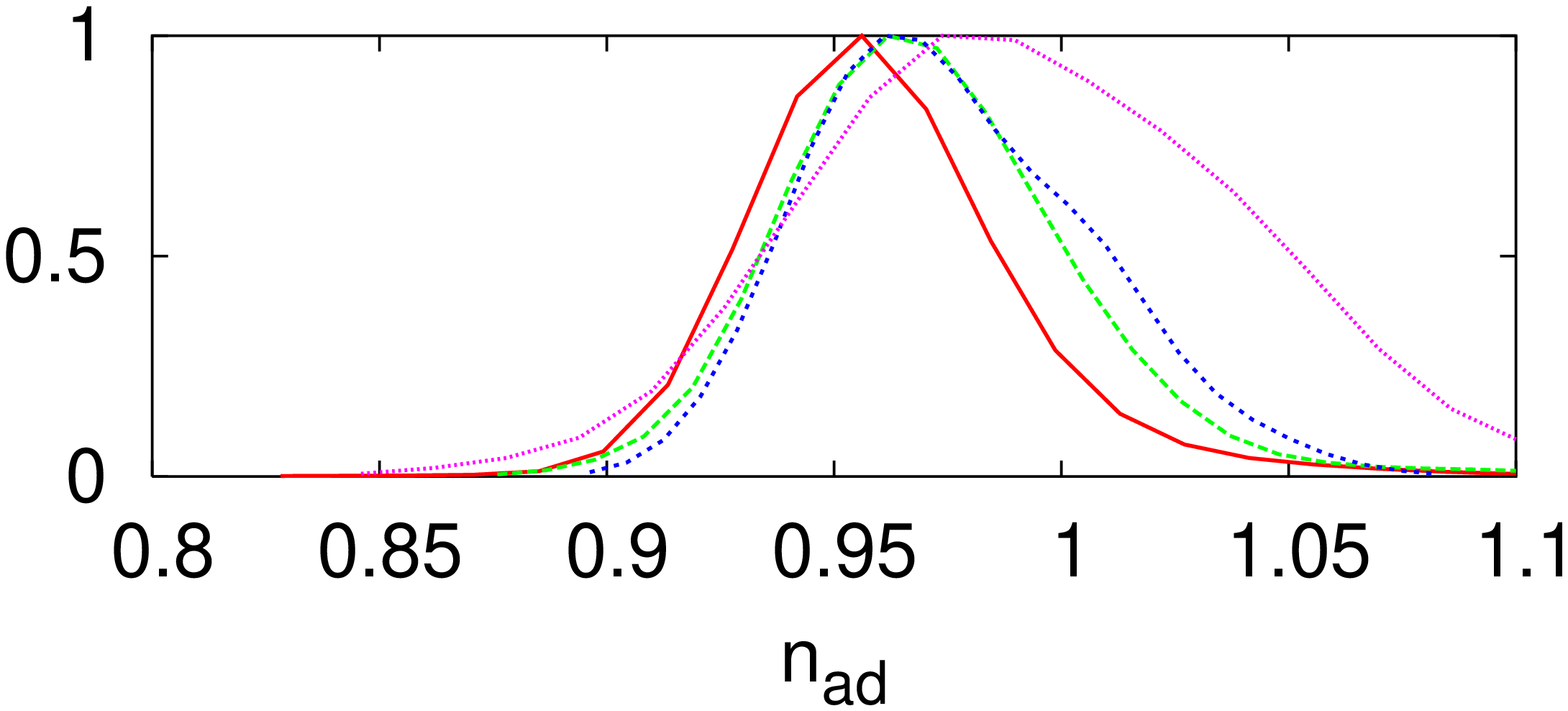}\\
\includegraphics[angle=-90,width=8.5cm]{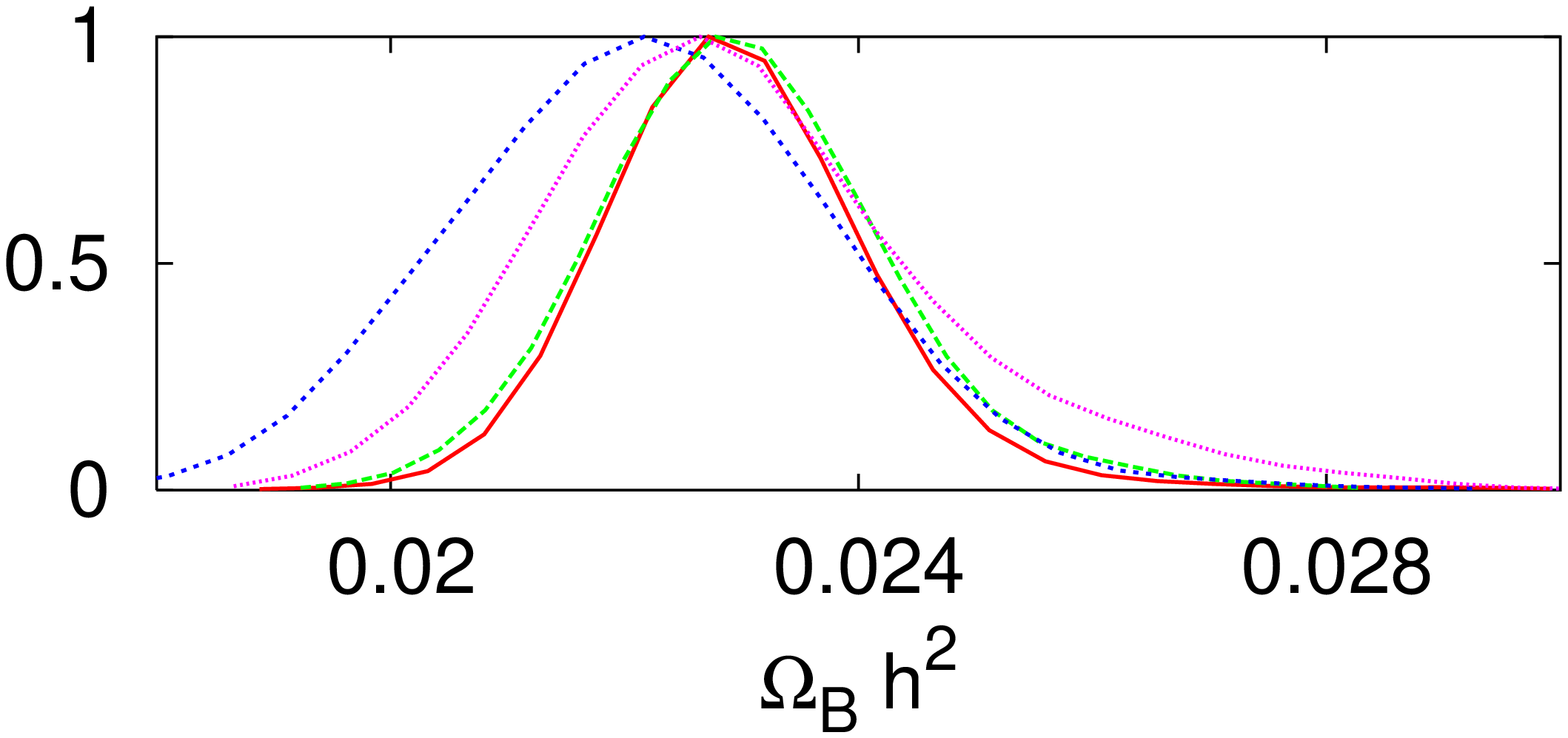}
\includegraphics[angle=-90,width=8.5cm]{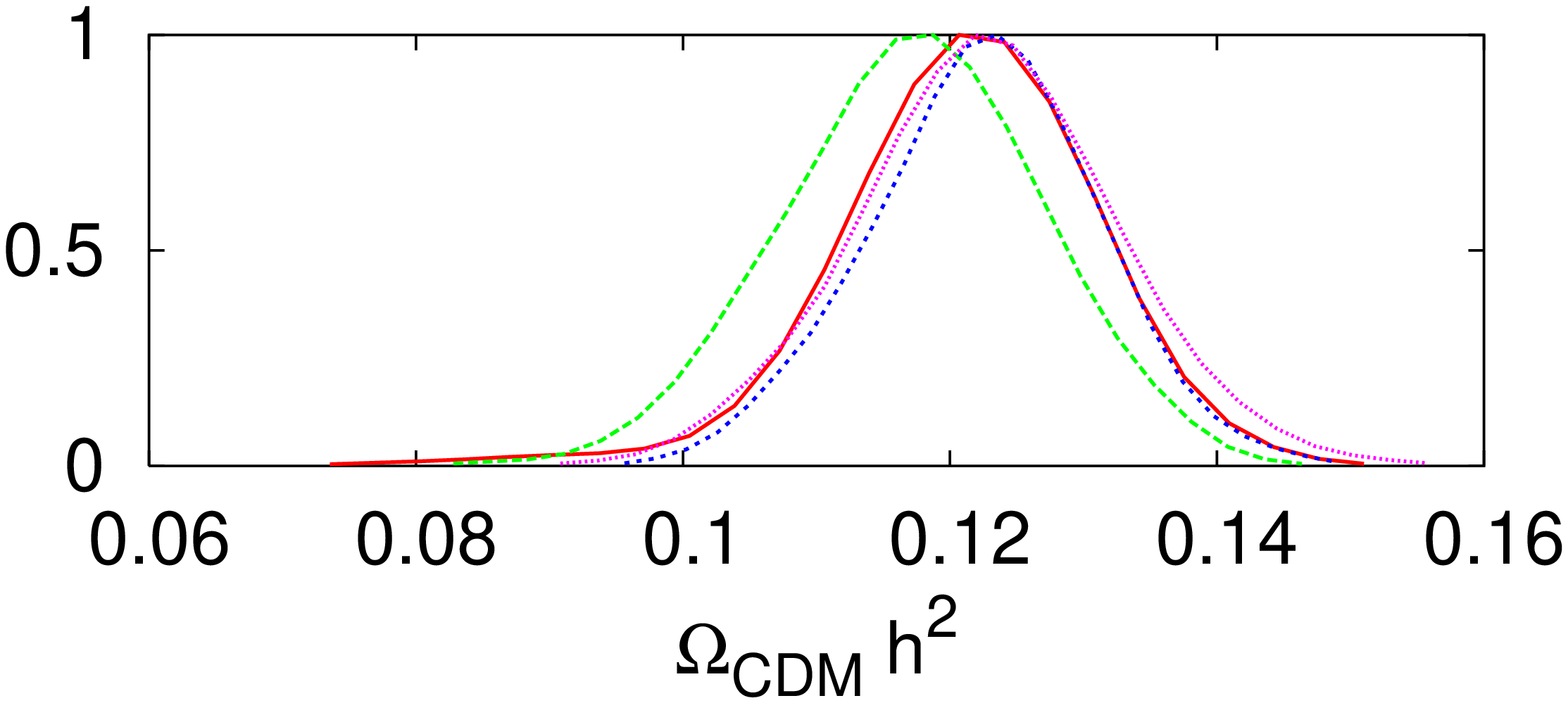}\\
\includegraphics[angle=-90,width=8.5cm]{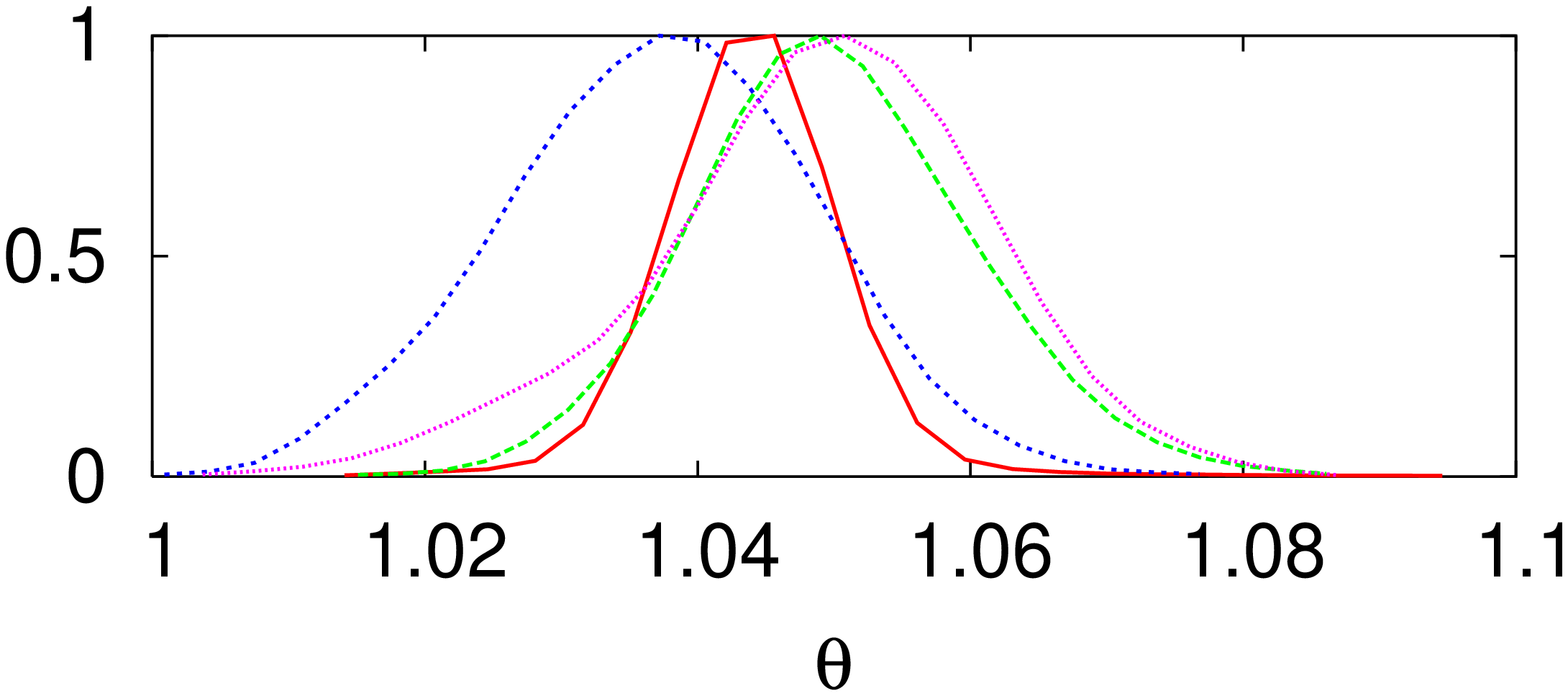}
\includegraphics[angle=-90,width=8.5cm]{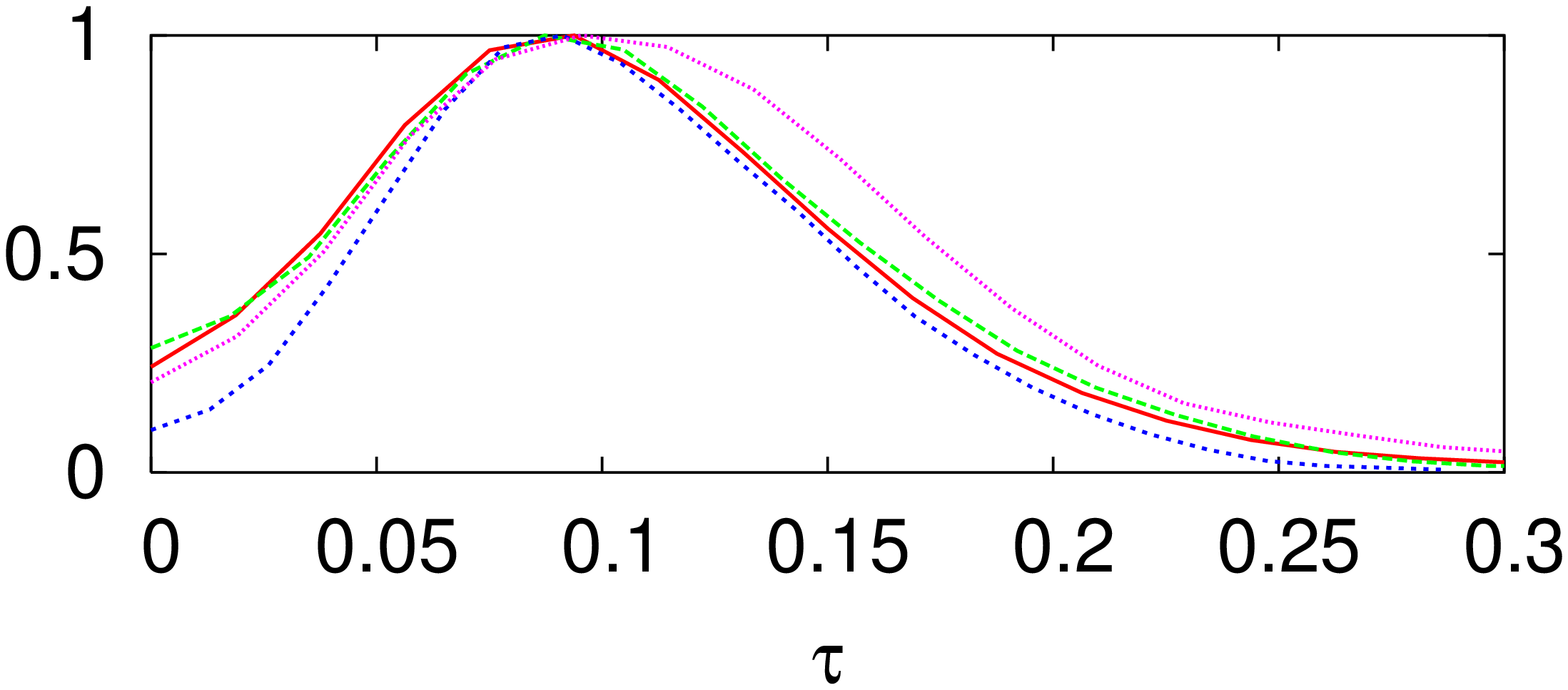}\\
\includegraphics[angle=-90,width=8.5cm]{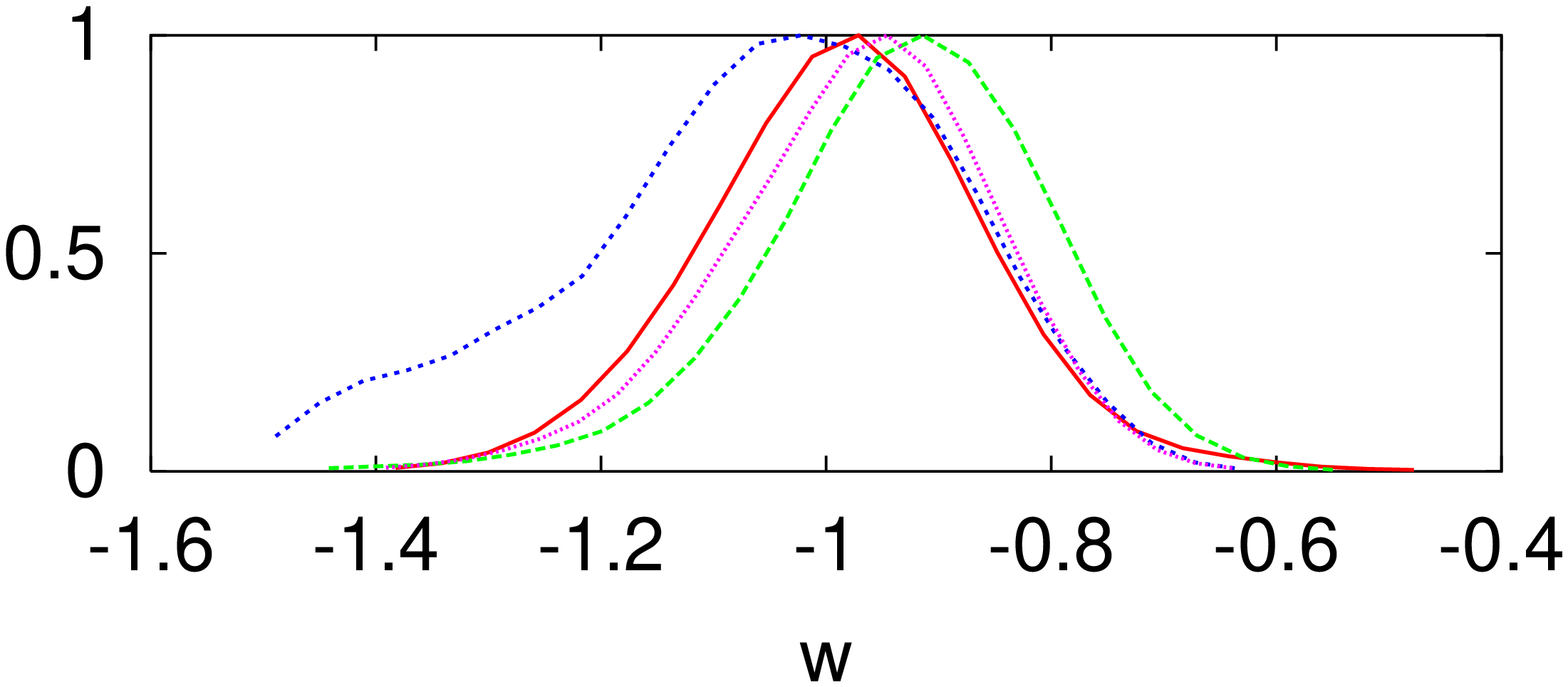}
\includegraphics[angle=-90,width=8.5cm]{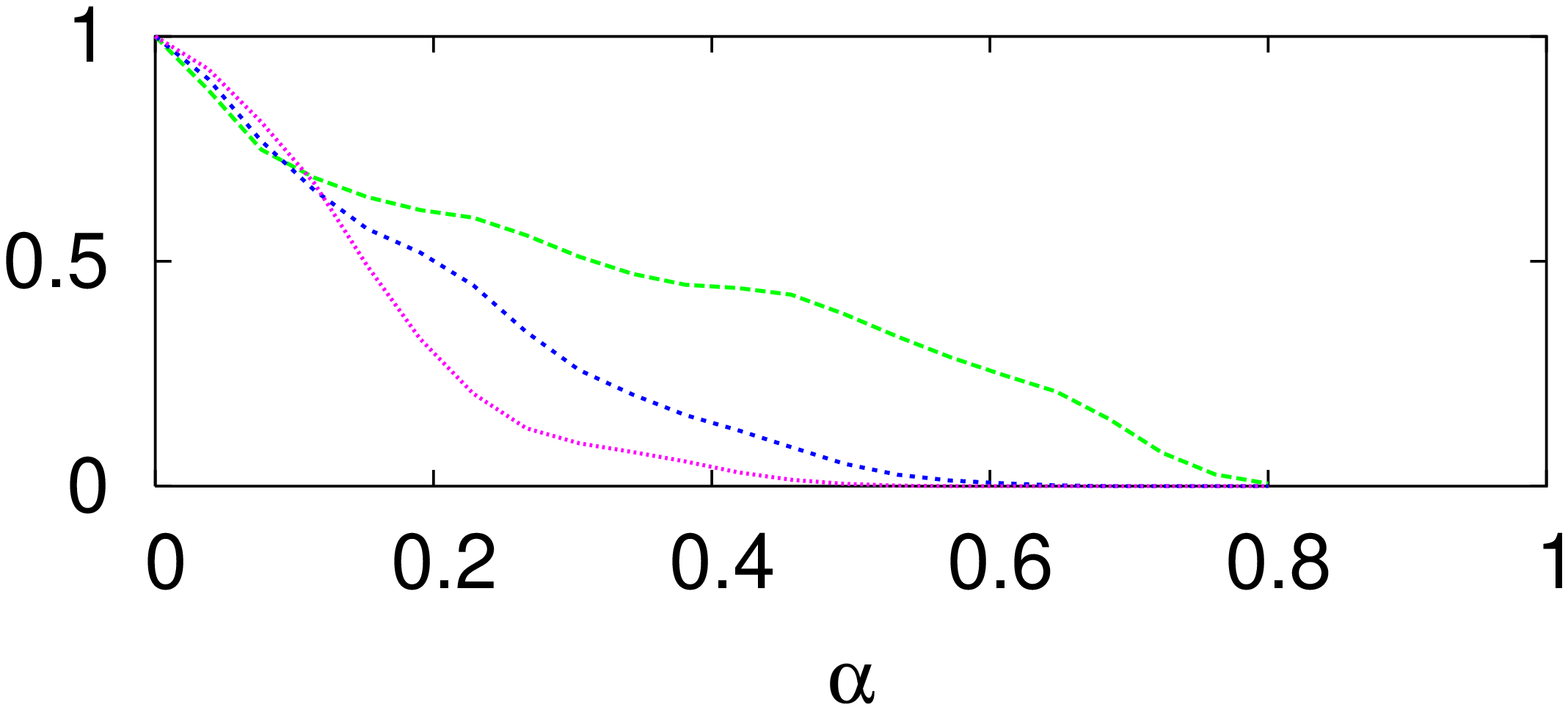}\\
\includegraphics[angle=-90,width=8.5cm]{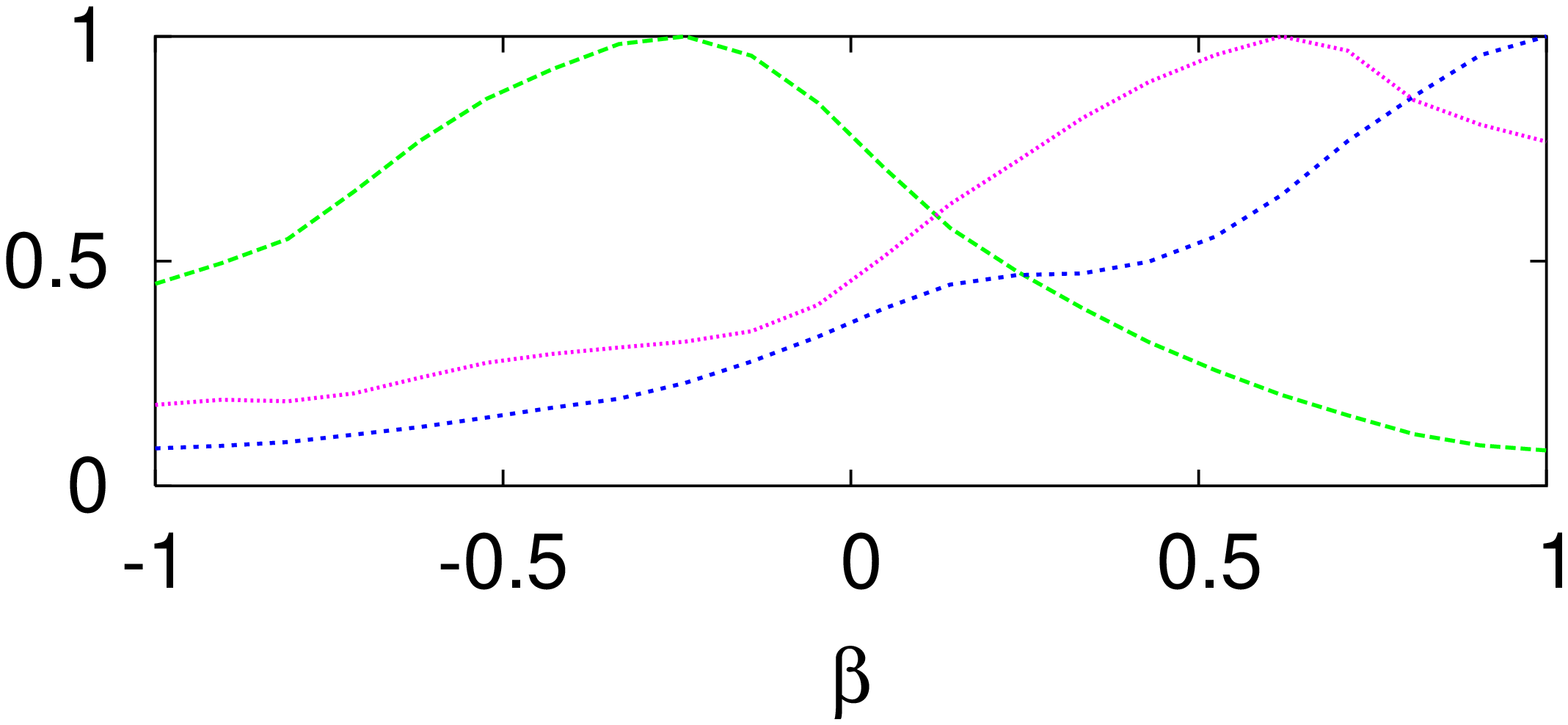}
\includegraphics[angle=-90,width=8.5cm]{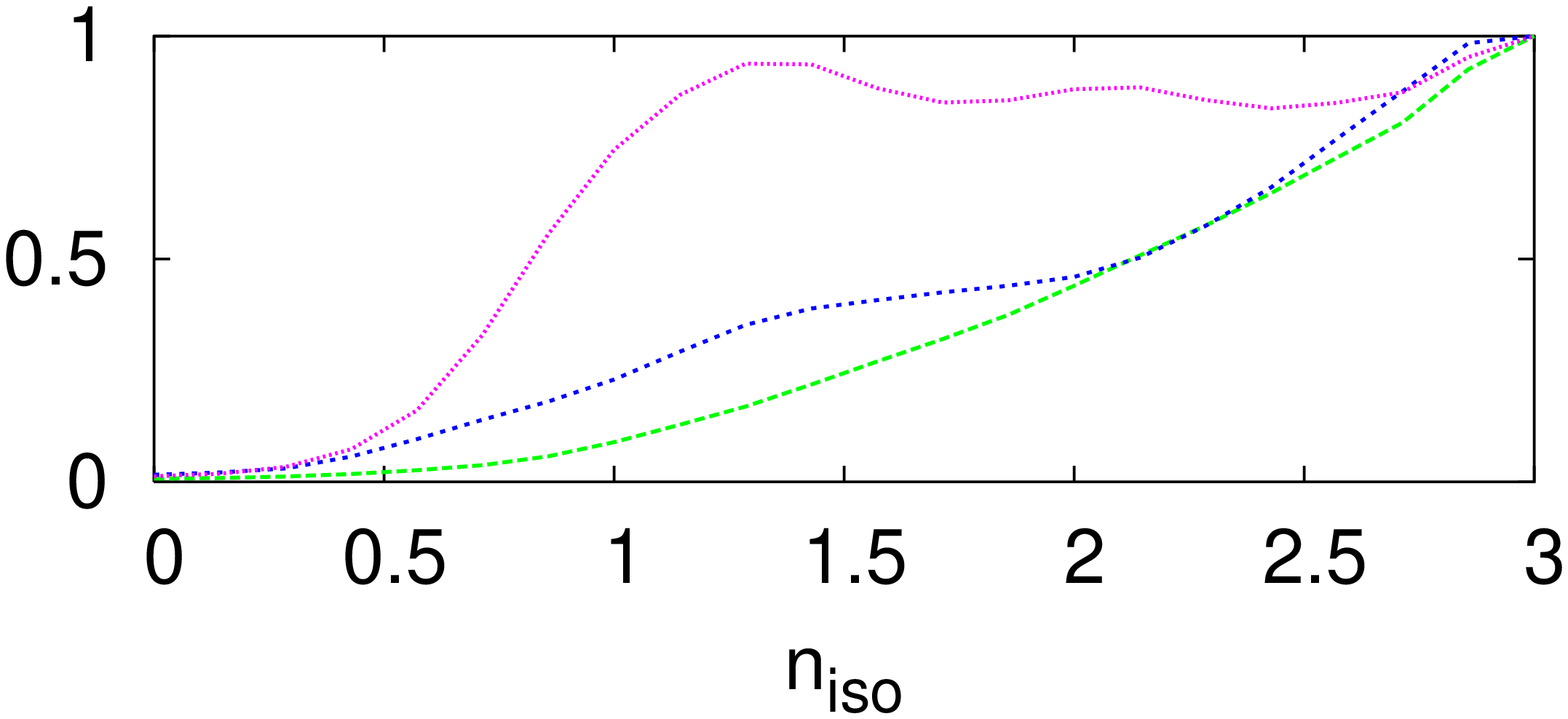}\\
\includegraphics[angle=-90,width=8.5cm]{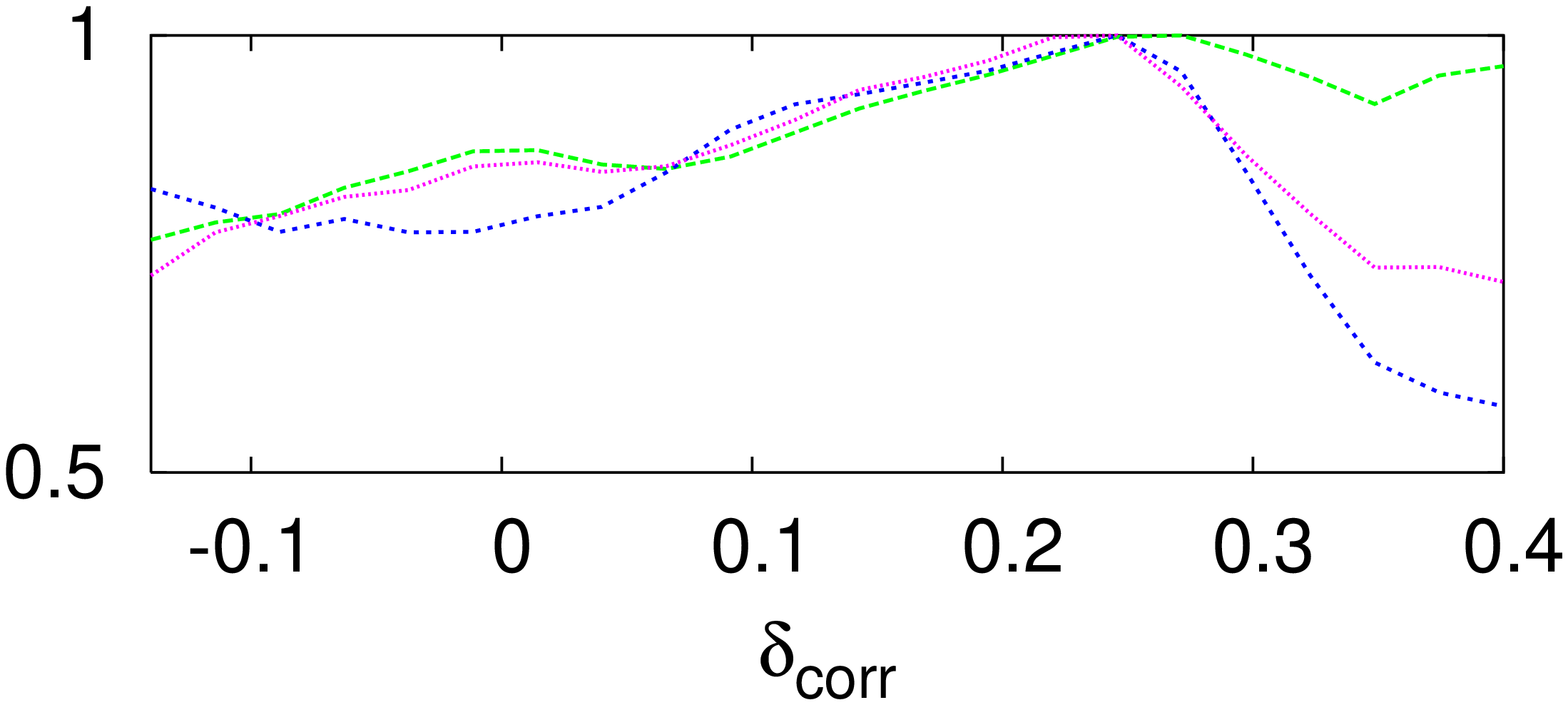}
\includegraphics[angle=-90,width=8.5cm]{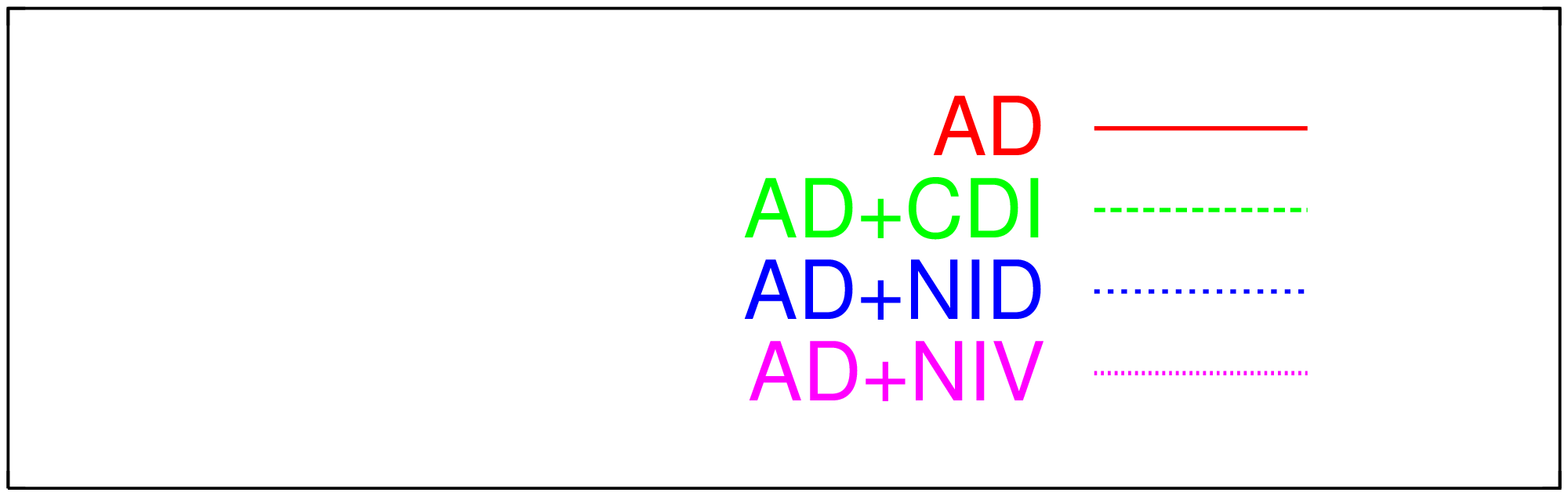}\\
\end{center}
\caption{\label{1Dplots} The one-dimensional likelihood functions for
our basis of eleven independent cosmological parameters (not including
the tilts of the two redshift surveys), for the adiabatic mode alone
(AD) or mixed with the three different types of isocurvature modes
(AD+CDI, AD+NID, AD+NIV). The first seven parameters are those of the
standard $\Lambda$CDM model, extended to dark energy with a constant
equation of state.  The last four parameters $(\alpha, \beta, n_{\rm
iso}, \delta_{\rm cor})$ describe the isocurvature initial
conditions. ($\delta_{\rm cor}$ is define in
Eq.~(\ref{eqdeltancor}.))}
\end{figure*}

\begin{figure*}[ht]
\begin{center}
\includegraphics[angle=-90,width=8.5cm]{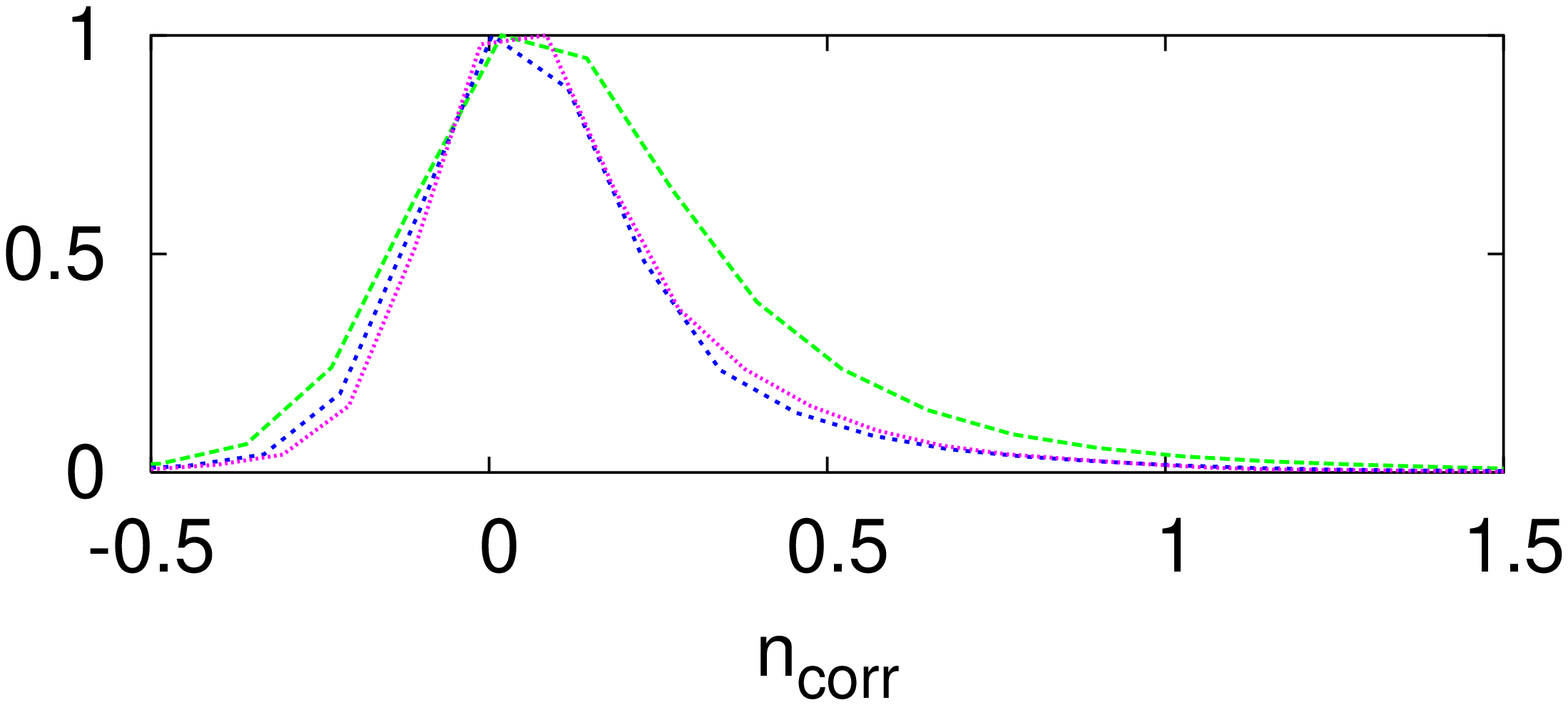}
\includegraphics[angle=-90,width=8.5cm]{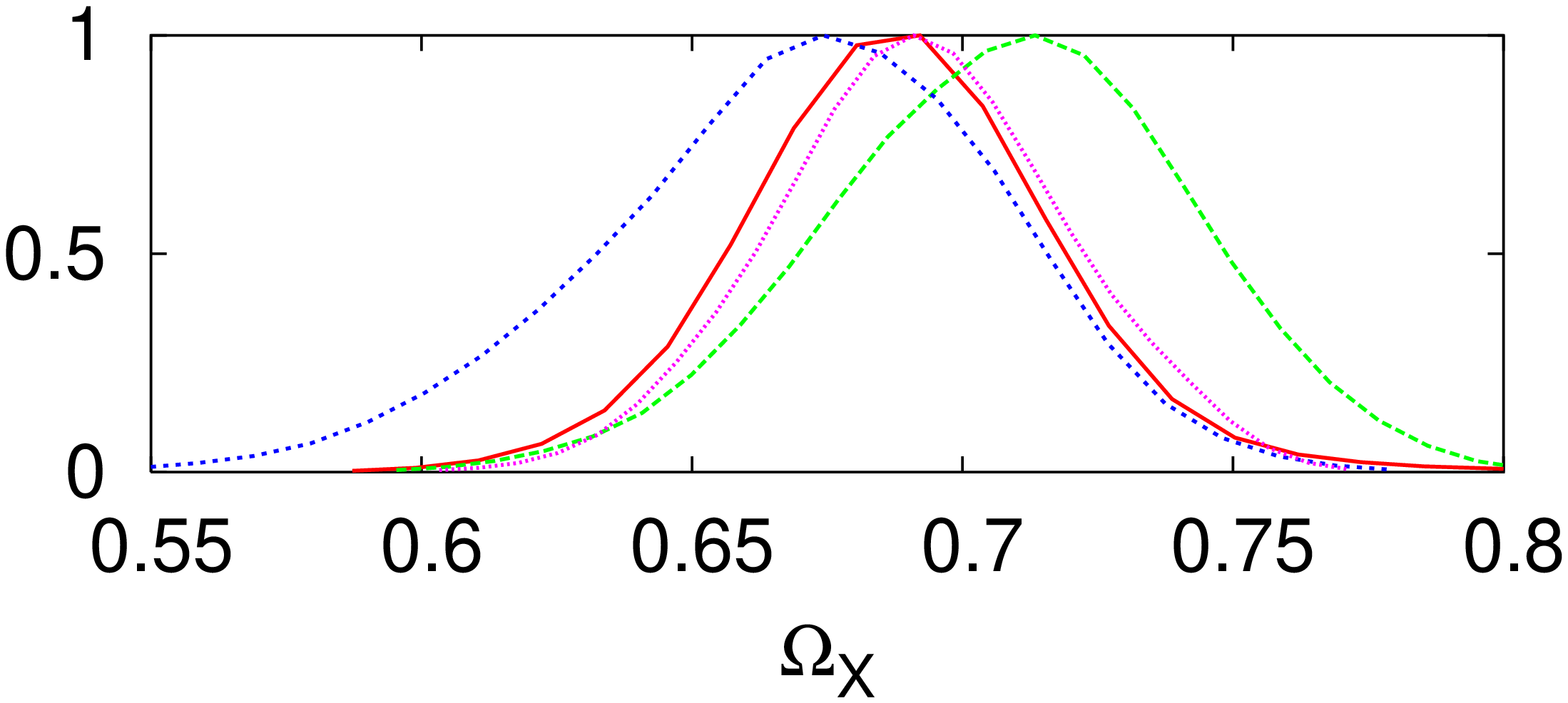}\\
\includegraphics[angle=-90,width=8.5cm]{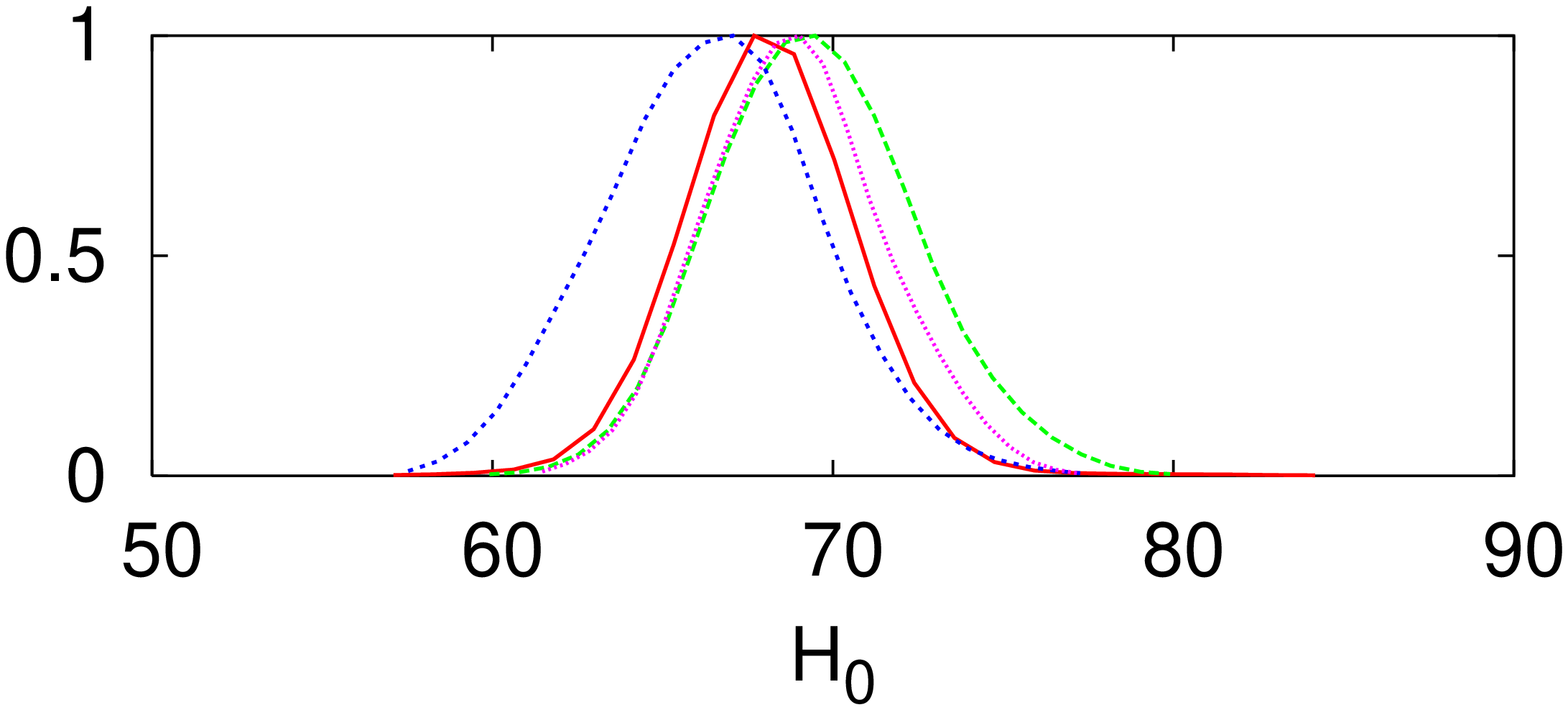}
\includegraphics[angle=-90,width=8.5cm]{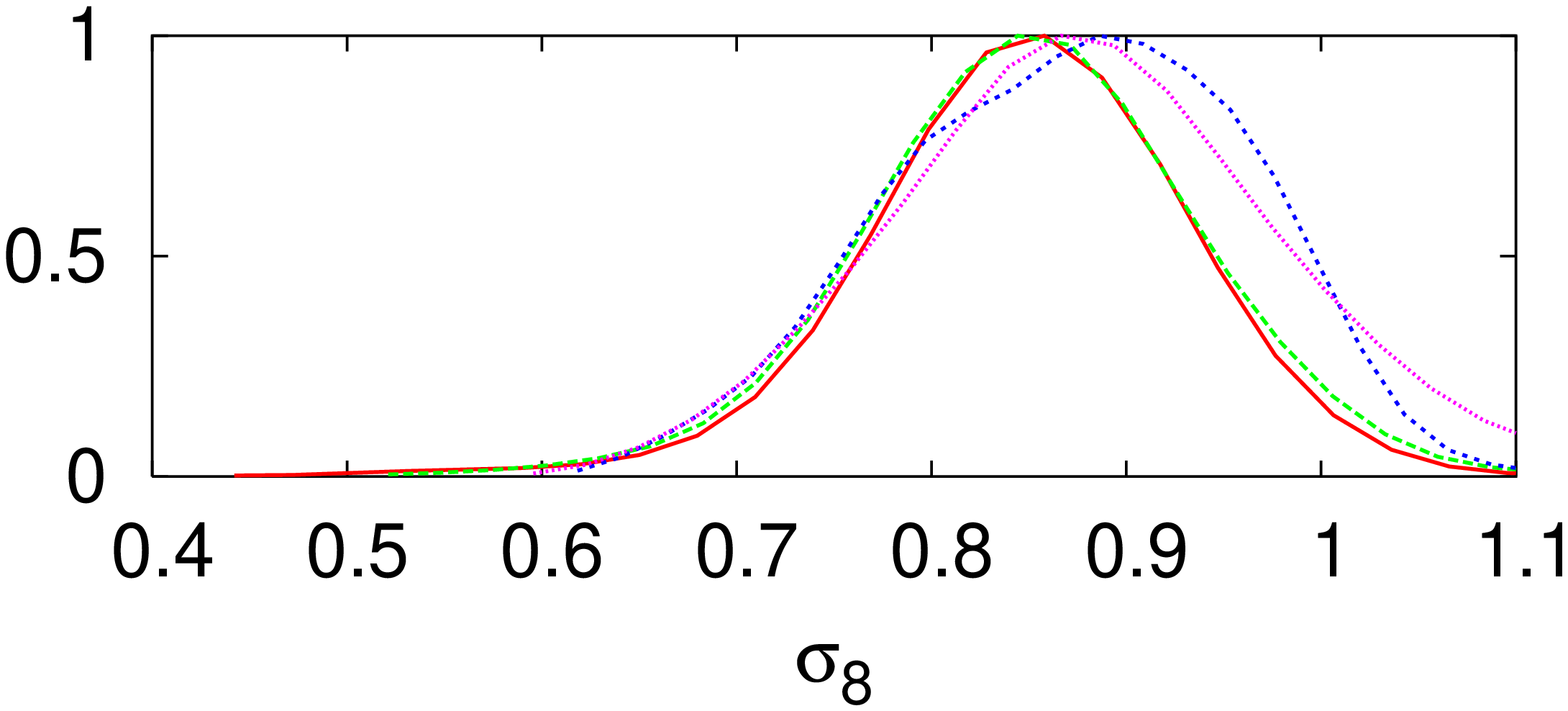}\\
\end{center}
\caption{\label{1Dplots_bis} Continuation of figure \ref{1Dplots},
showing the 1D likelihood of some derived cosmological parameters, for
the same cases. These likelihoods should be considered with care,
because the parameters shown here do not belong to the basis used by
the Markov chain algorithm. Therefore, the shape of the above
likelihoods depends not only on the likelihood of the underlying parameters,
but also on the properties of the functions relating them to the parameters
of the basis. This explains for instance 
why $n_{\rm cor}=\delta_{\rm cor} \, \ln(|\beta|^{-1})$ 
seems to be well-constrained, while
$\delta_{\rm cor}$ and $\beta$ are not.}
\end{figure*}

\begin{figure*}[ht]
\includegraphics[angle=-90,width=8.7cm]{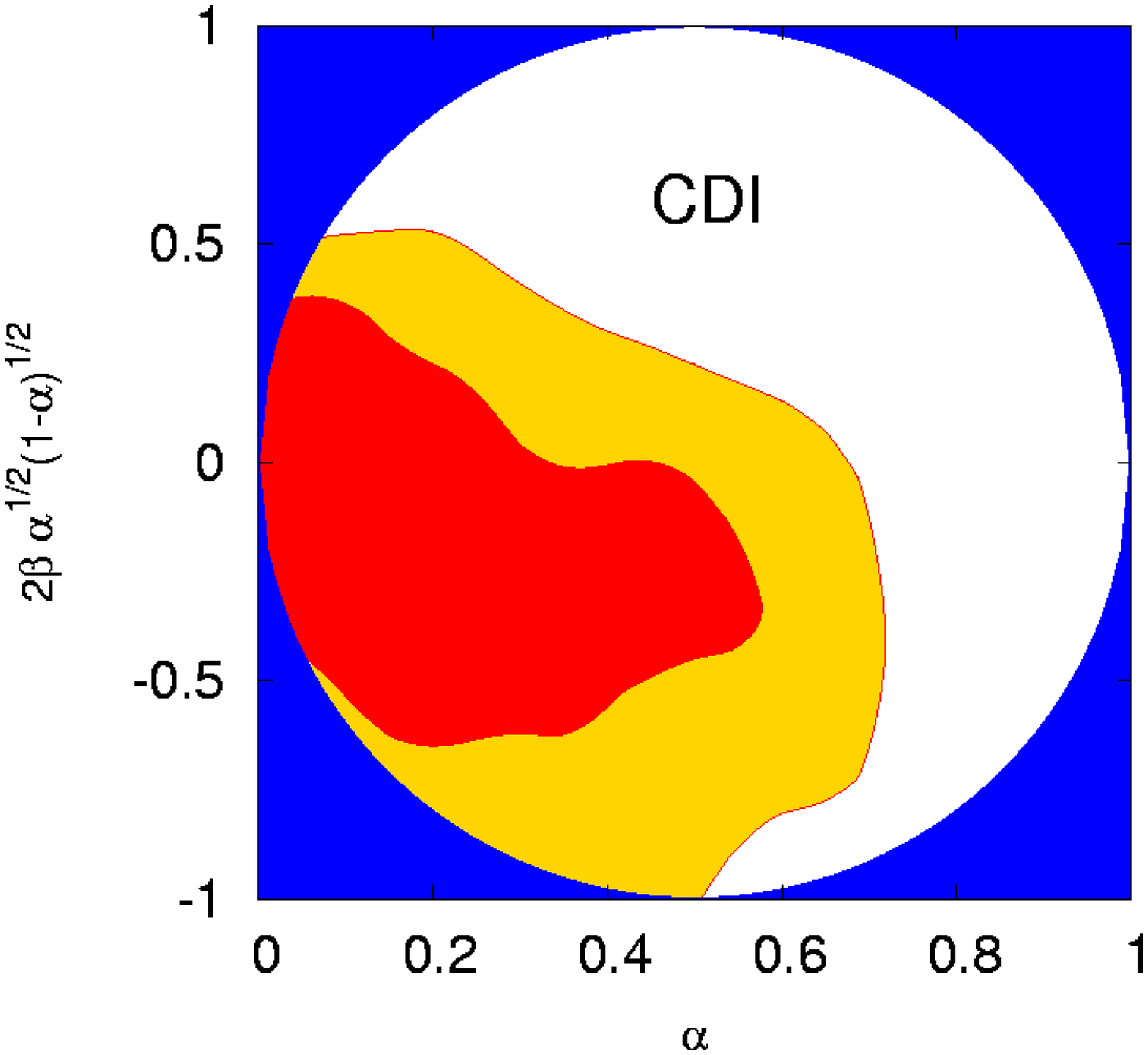}
\includegraphics[angle=-90,width=8.7cm]{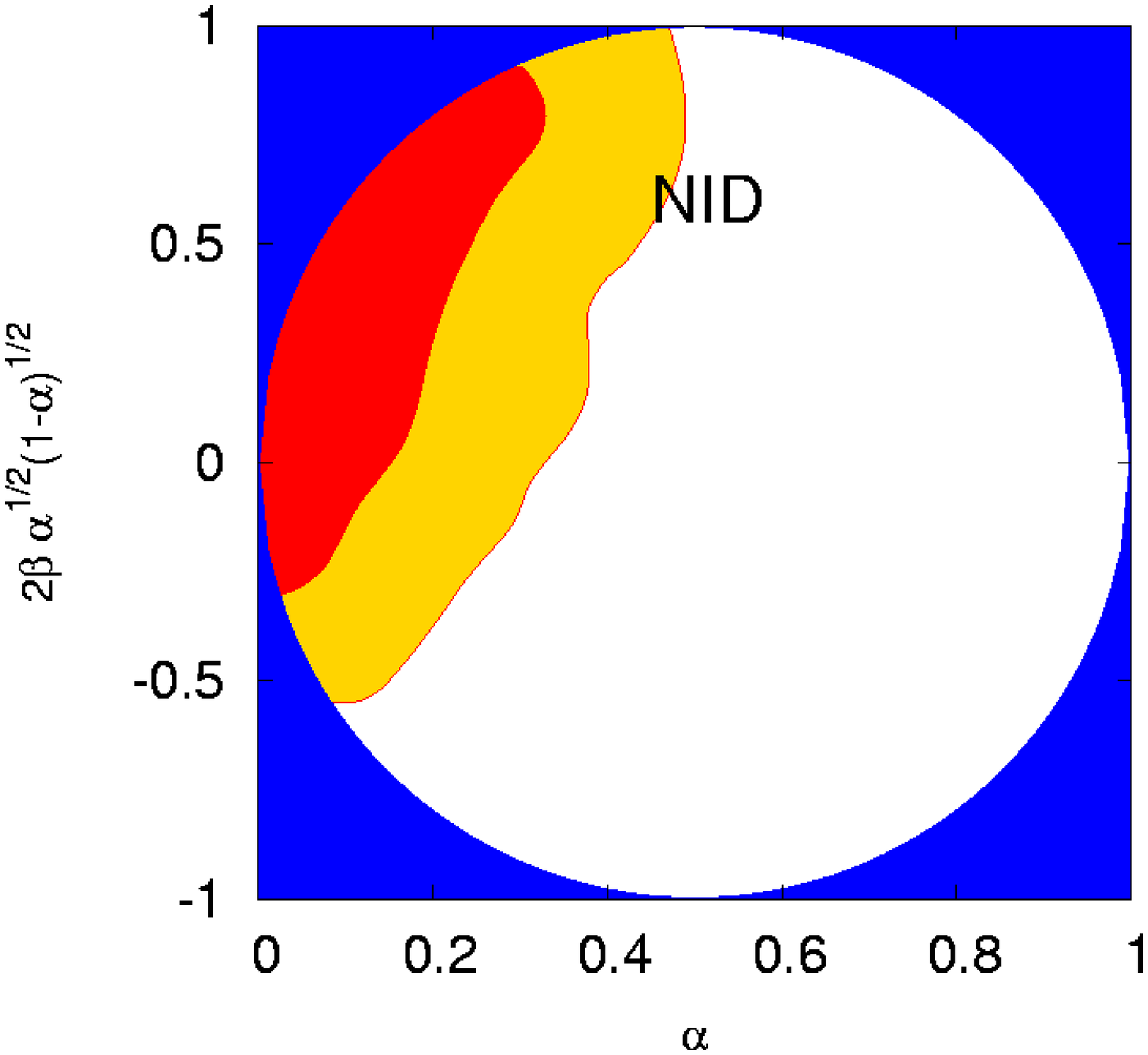}\\
\includegraphics[angle=-90,width=8.7cm]{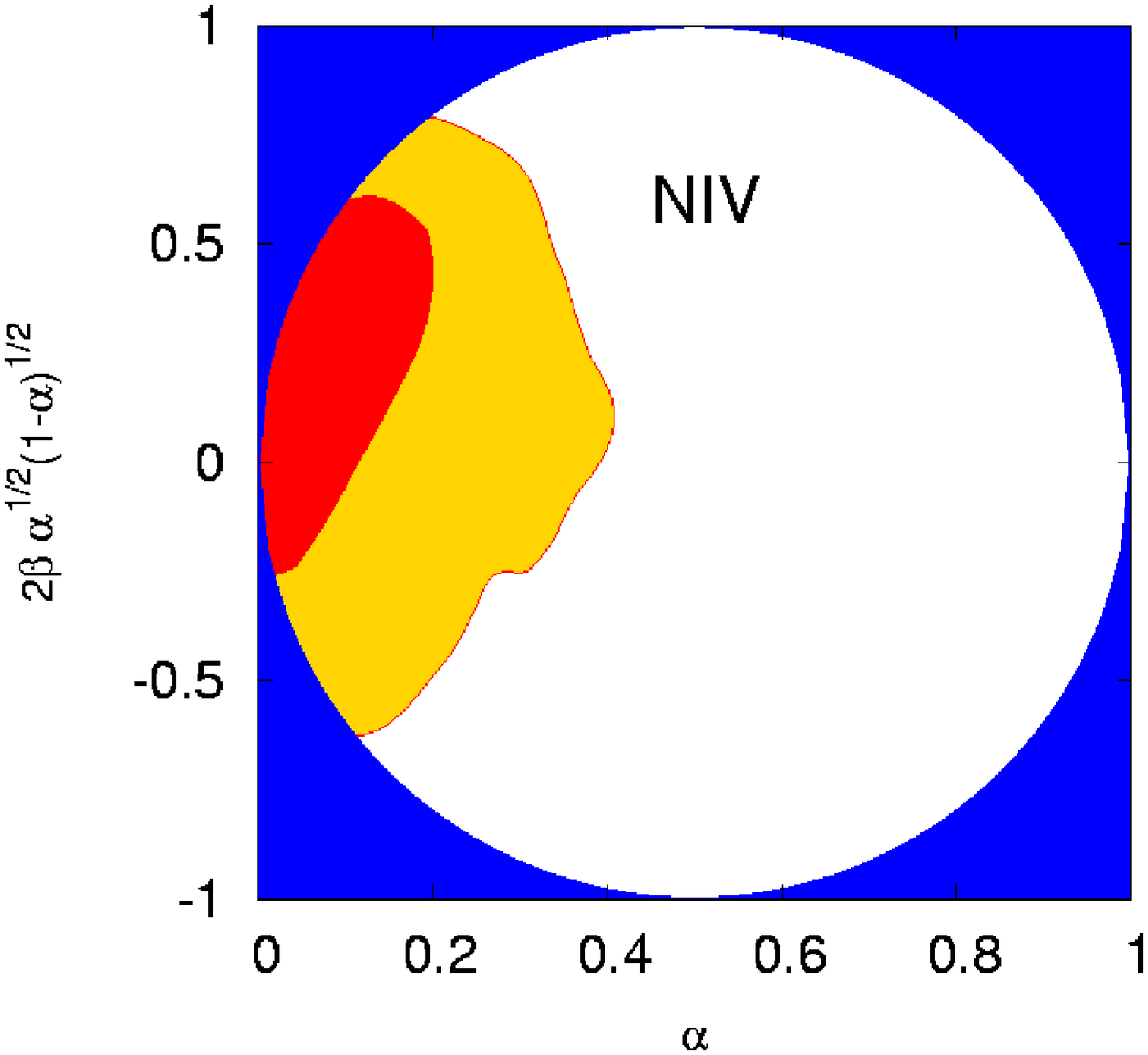}
\includegraphics[angle=-90,width=8.7cm]{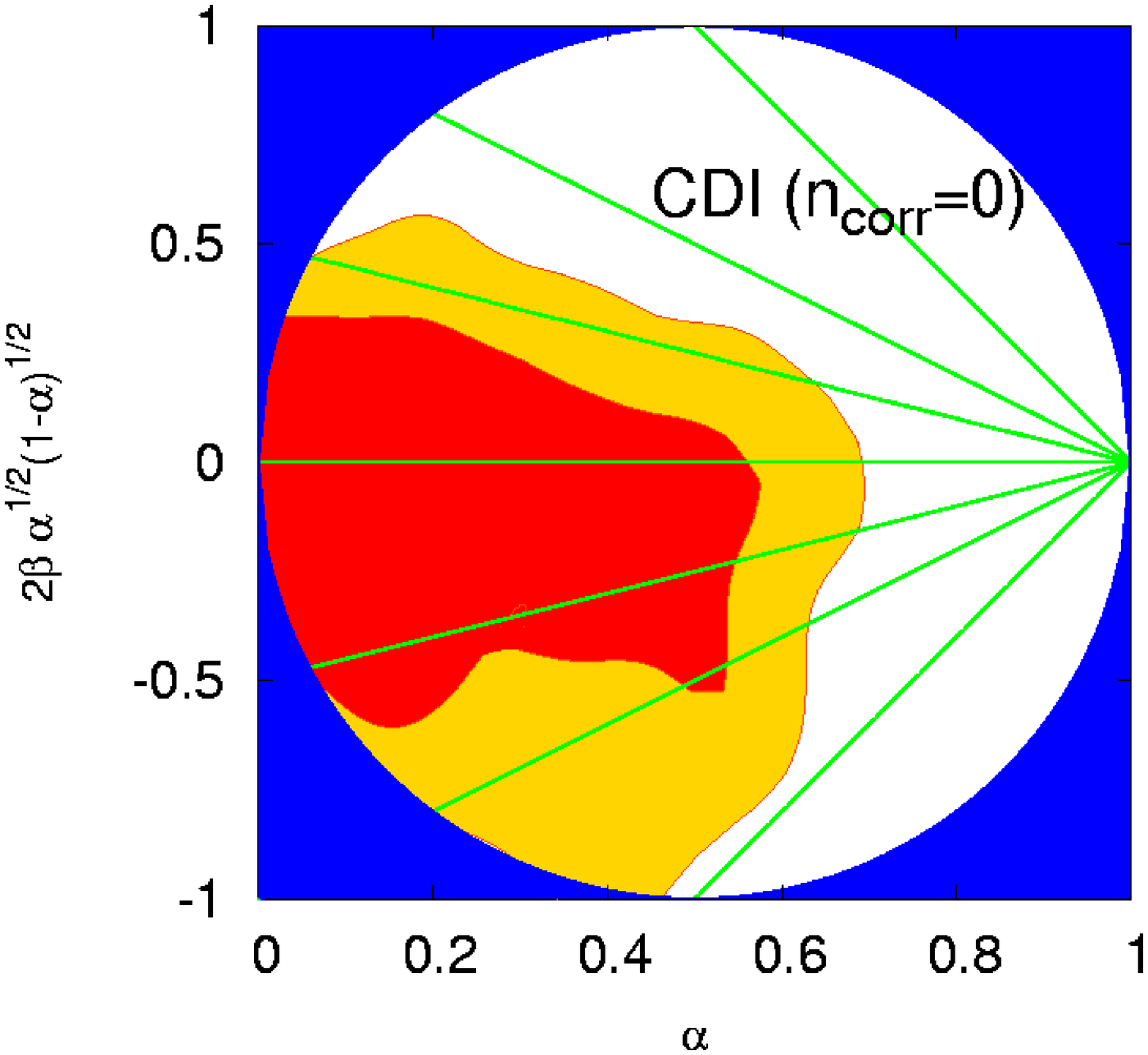}
\caption{\label{fig:alpha} The 2-$\sigma$ contours of $\alpha$ 
and the cross-correlated mode
coefficient $2 \beta\sqrt{\alpha(1-\alpha)}$, for a) the CDI
isocurvature mode; b) the NID mode; c) the NIV mode; d) the CDI mode,
with the constraint $n_{\rm cor}=0$ and the contours of equal
$2(R^2-1)/s_k=0, \pm 0.5, \pm 1, \pm 2$ from double inflation.
}
\end{figure*}

In order to take into account the $\beta$-dependent constraint on
$n_{\rm cor}$, see eq.~(\ref{ncorrconst}), we choose to define
a new parameter 
\begin{equation}
\label{eqdeltancor}
\delta_{\rm cor} \equiv n_{\rm cor} / \ln(|\beta|^{-1}),
\end{equation}
whose boundaries are fixed once and for all by the values of the pivot
scale and the scales ($k_{\rm min}$, $k_{\rm max}$) introduced in the
previous section. The basis of cosmological parameters used by the
Markov Chain algorithms consists on: (i) the seven parameters
describing the standard adiabatic $\Lambda$CDM model, extended to dark
energy with a constant equation of state, and (ii) the four parameters
describing the admixture of one (correlated) isocurvature mode,
already defined in the previous section. More explicitely, we use the
following basis: 1) the overall normalisation parameter $\ln[10^{10}
{\cal R}_{\rm rad}]$ where ${\cal R}_{\rm rad}$ is the curvature
perturbation in the radiation era; 2) the adiabatic tilt $\nad$; 3)
the baryon density $\oB=\OB h^2$; 4) the cold dark matter density
$\ocdm=\Ocdm h^2$; 5) the ratio $\theta$ of the sound horizon to the
angular diameter distance multiplied by 100; 6) the optical depth to
reionization $\tau$; 7) the dark energy equation of state parameter
$w$; then, for the isocurvature sector; 8) the isocurvature
contribution $\alpha$; 9) the cross-correlation parameter $\beta$;
10) the isocurvature tilt $n_{\rm iso}$; 11) the parameter
$\delta_{\rm cor}$ which determines the cross-correlation tilt
$\ncor$; and finally, two arbitrary bias parameters associated to the
2dF and SDSS power spectrum.  Our full parameter space is therefore
13-dimensional.

We did not devote a specific analysis to the case of the baryon
isocurvature modes, which is qualitatively similar to that of CDI modes,
since the spectra are simply rescaled by a factor $\Omega_{\rm
B}^2/\Omega_{\rm cdm}^2$ ($\Omega_{\rm B}/\Omega_{\rm cdm}$ for the
cross-correlation): thus, compared to the AD + CDI case, significantly
larger values of $\alpha$ will be allowed in the AD + BI case. Like in
other recent analyses, we find that the inclusion of isocurvature modes
does not improve significantly the goodness-of-fit of the cosmological
model, since in the AD+CDI, AD+NID and AD+NIV cases the minimum $\chi^2$
is always between 1672 and 1674 for 1614 degrees of freedom, to be
compared with 1674 for 1618 degrees of freedom in the pure adiabatic
case. Therefore, the question is just to study how much departure from
the standard picture is allowed, by computing the Bayesian confidence
limit on the isocurvature parameters. A more detailed analysis of model
comparison with Bayesian Information Criteria~\cite{Liddle:2004nh} will
be done in a follow up paper.

On Fig.~\ref{1Dplots} we plot the marginalised likelihood for our
basis of eleven cosmological parameters, in the cases AD+CDI, AD+NID
and AD+NIV, compared with the pure adiabatic case.
Figure~\ref{1Dplots_bis} shows the likelihood of some derived
parameters. It appears that most parameters are robust against the
inclusion of isocurvature perturbations (this is in agreement with the
conclusion of Ref.~\cite{Moodley:2004ws} that with only one
isocurvature mode present, no significant parameter degeneracy pops
out). Our 95\% C.L. on $\alpha$ in the three cases is given in
Table~\ref{alpha_bounds}.
\begin{table}[h]
\caption{The one-dimensional 2-$\sigma$ ranges on 
the isocurvature mode coefficients for the various models.}
\begin{tabular}{|c| c c |}
\hline
model &
$\alpha$ & \hspace{2mm} $2 \beta\, [\alpha (1 - \alpha)]^{1/2}$\cr 
\hline \hline
AD+CDI & $< 0.6$ &\hspace{2mm} 
$-0.7$ \ to \ $0.3$ \cr
AD+NID & 
$< 0.4$ &\hspace{2mm}  
$-0.2$ \ to \ $0.8$ \cr
AD+NIV & $< 0.3$ & \hspace{2mm} 
$-0.4$ \ to \ $0.6$ \cr
\hline
\end{tabular}
\label{alpha_bounds}
\end{table}

Note that in the limit $\alpha=0$, the three parameters $\beta$,
$\niso$, $\delta_{\rm cor}$ become irrelevant. So, the fact that pure
adiabatic models are very good fits implies that these parameters are
loosely constrained. This explains why the corresponding likelihoods
on Fig.~\ref{1Dplots} are not well-peaked like for other
parameters. In addition, these likelihoods should be considered with
great care, because it is difficult for the Markhov Chains to explore
in detail the tails of the multi-dimensionnal likelihood
corresponding to tiny values of $\alpha$, where basically any value of
($\beta$, $\niso$, $\delta_{\rm cor}$) are allowed. Therefore,
increasing the number of samples would tend to flatten these
likelihoods, while the other ones would remain stable (as we checked
explicitely). However, it is clear that all models prefer a large
isocurvature tilt and saturate the bound $\niso<3$
that we fixed in the present analysis. This feature is important for
understanding our results and comparing with other analyses, as
explained in the last paragraph of this section.

On Fig.~\ref{fig:alpha}, we plot the two-dimensional confidence levels
directly for the isocurvature and cross-correlation coefficients
($\alpha$, $2 \beta\sqrt{\alpha(1-\alpha)}$) in the three cases AD+CDI,
AD+NID and AD+NIV. The last plot corresponds to the AD+CDI case with a
prior $n_{\rm cor}=0$ which is relevant for the bounds on double
inflation, but the results are not substantially different from the
general AD+CDI case. We see that the AD+CDI model slightly prefers
anti-correlated cases (note that $\beta<0$ means a positive contribution
from the cross-correlated component), while AD+NID and AD+NIV models
clearly prefer correlated ones.

\begin{figure*}[ht]
\begin{center}
\includegraphics[angle=-90,width=8cm]{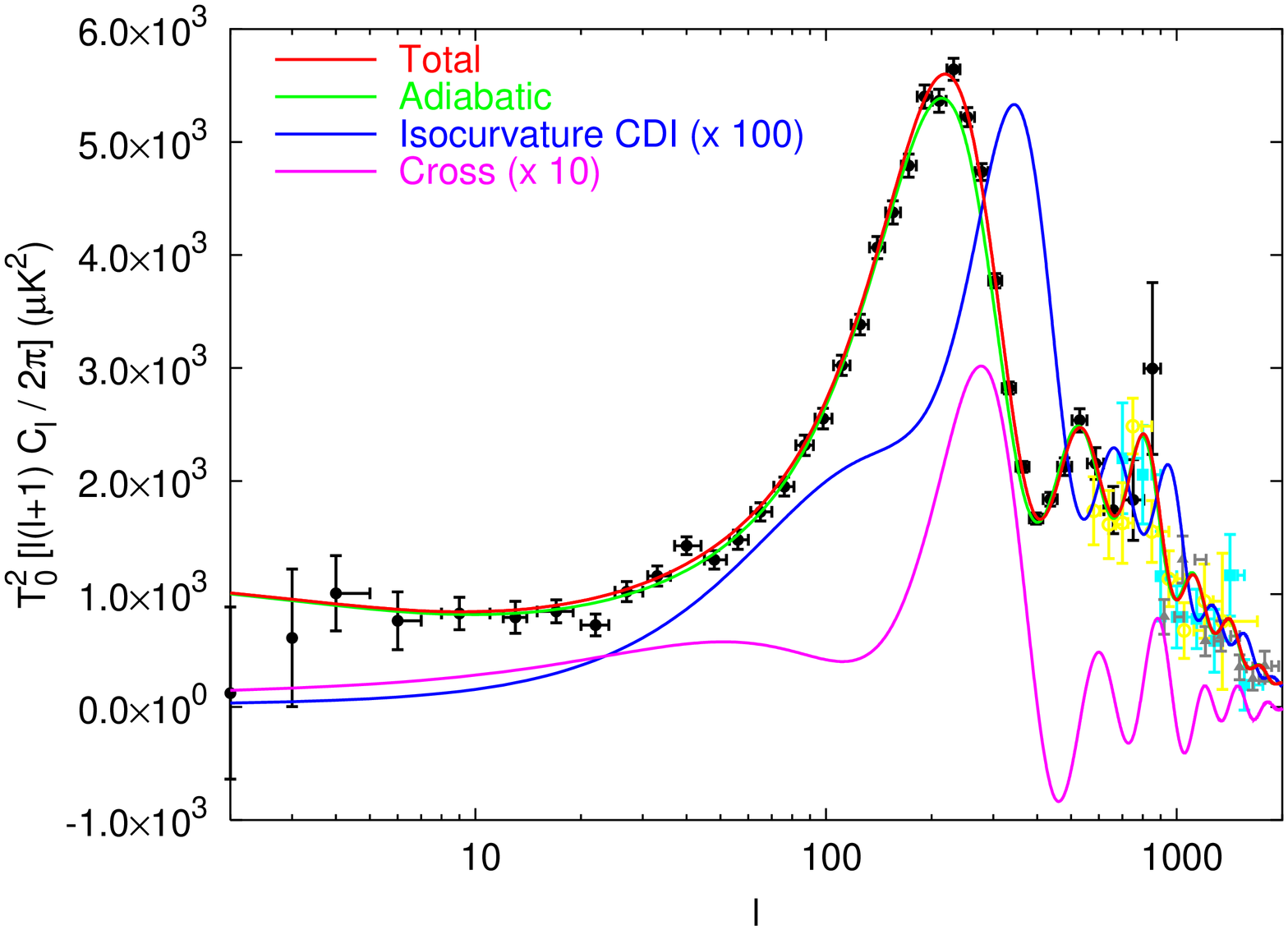}~
\includegraphics[angle=-90,width=8cm]{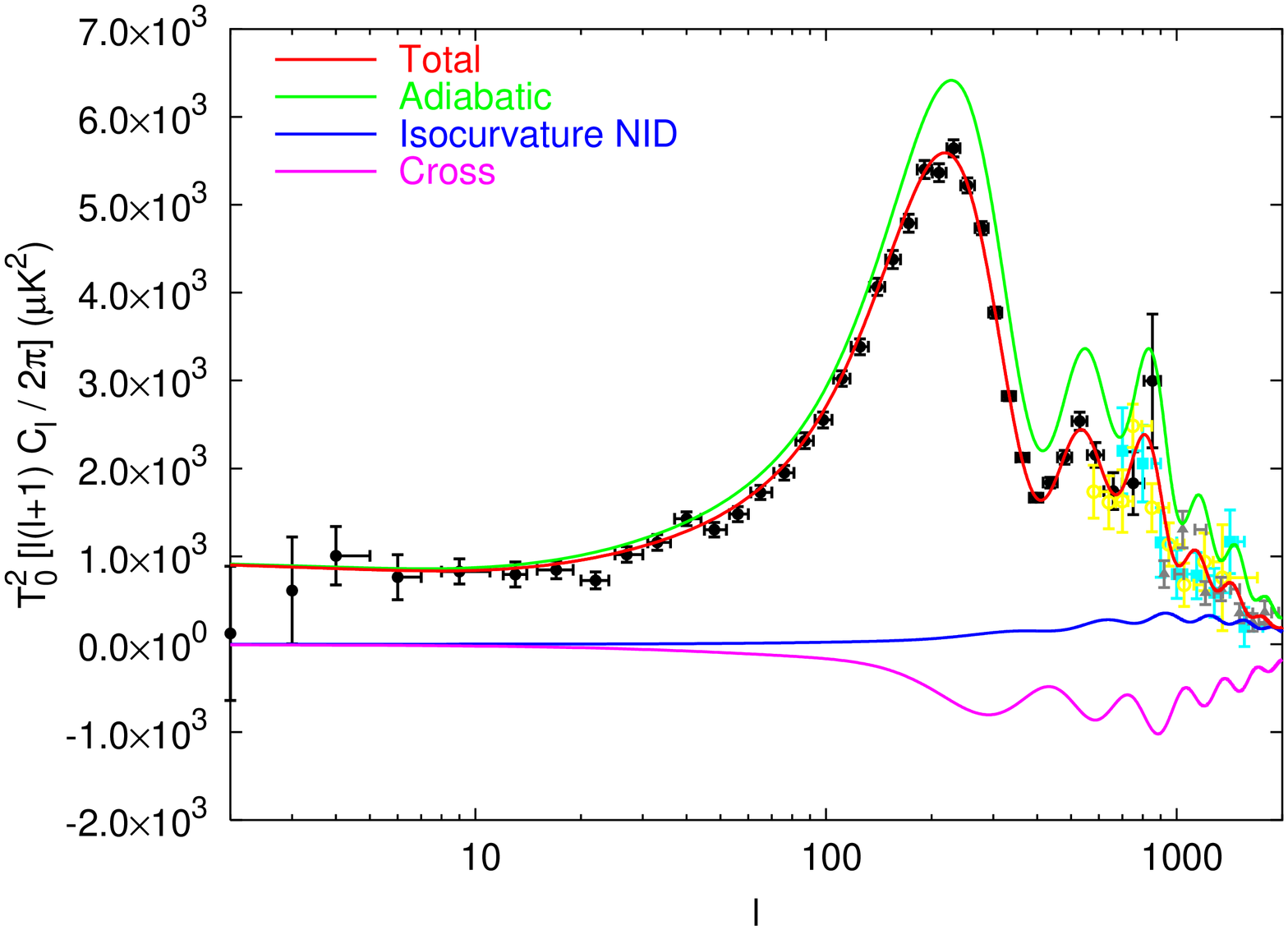}\\[2mm]
\includegraphics[angle=-90,width=8cm]{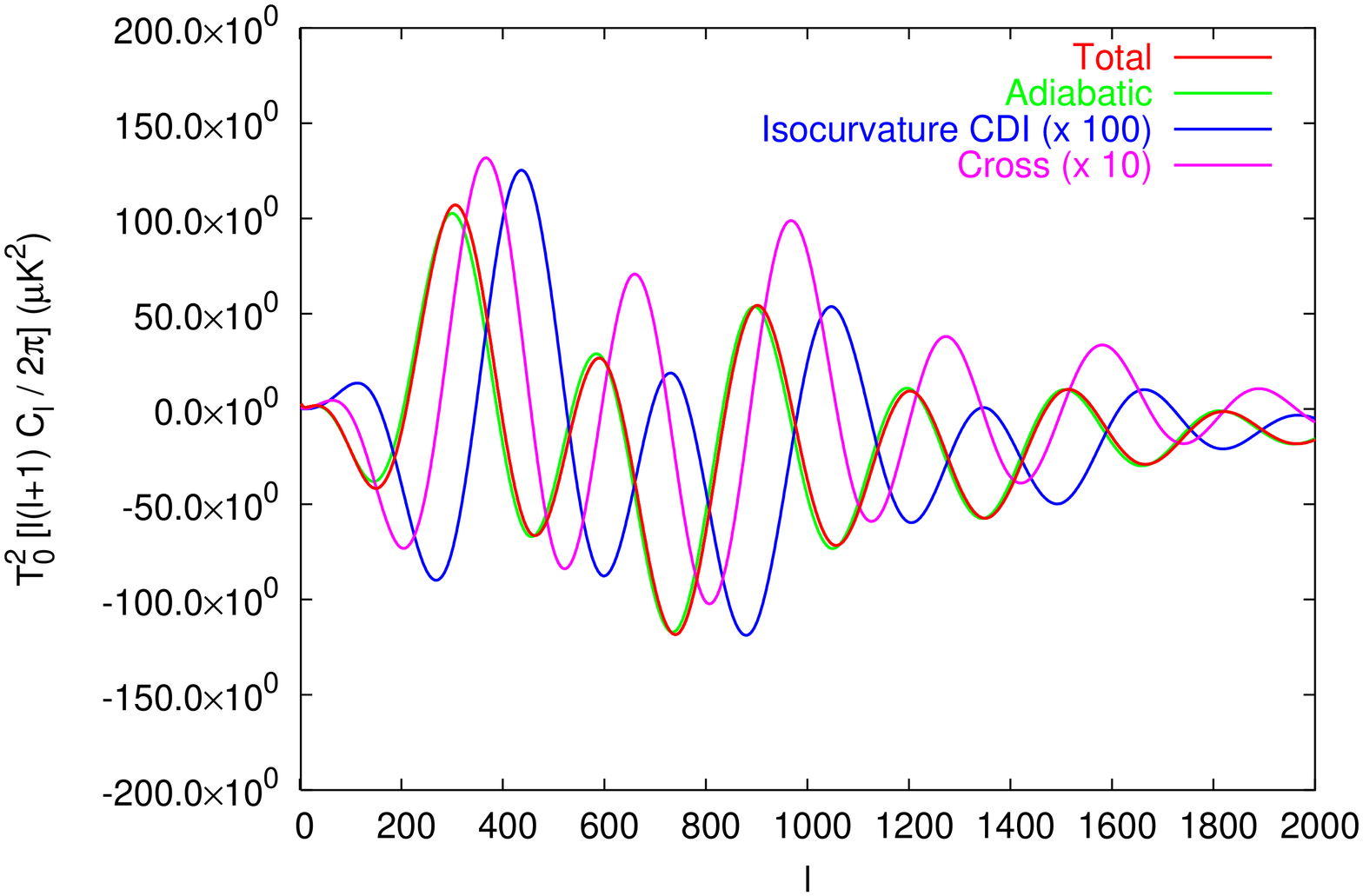}~
\includegraphics[angle=-90,width=8cm]{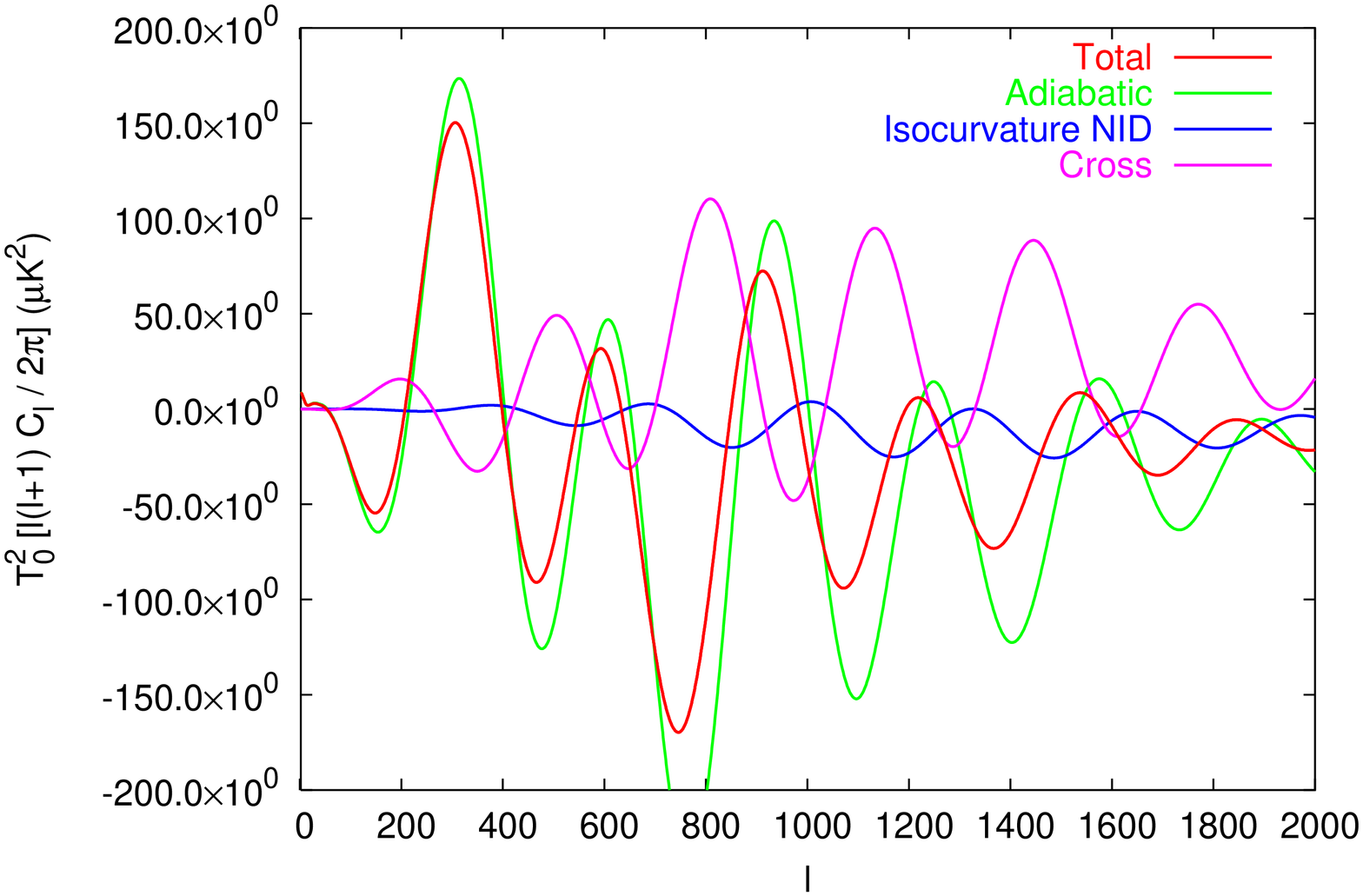}\\[2mm]
\includegraphics[angle=-90,width=8cm]{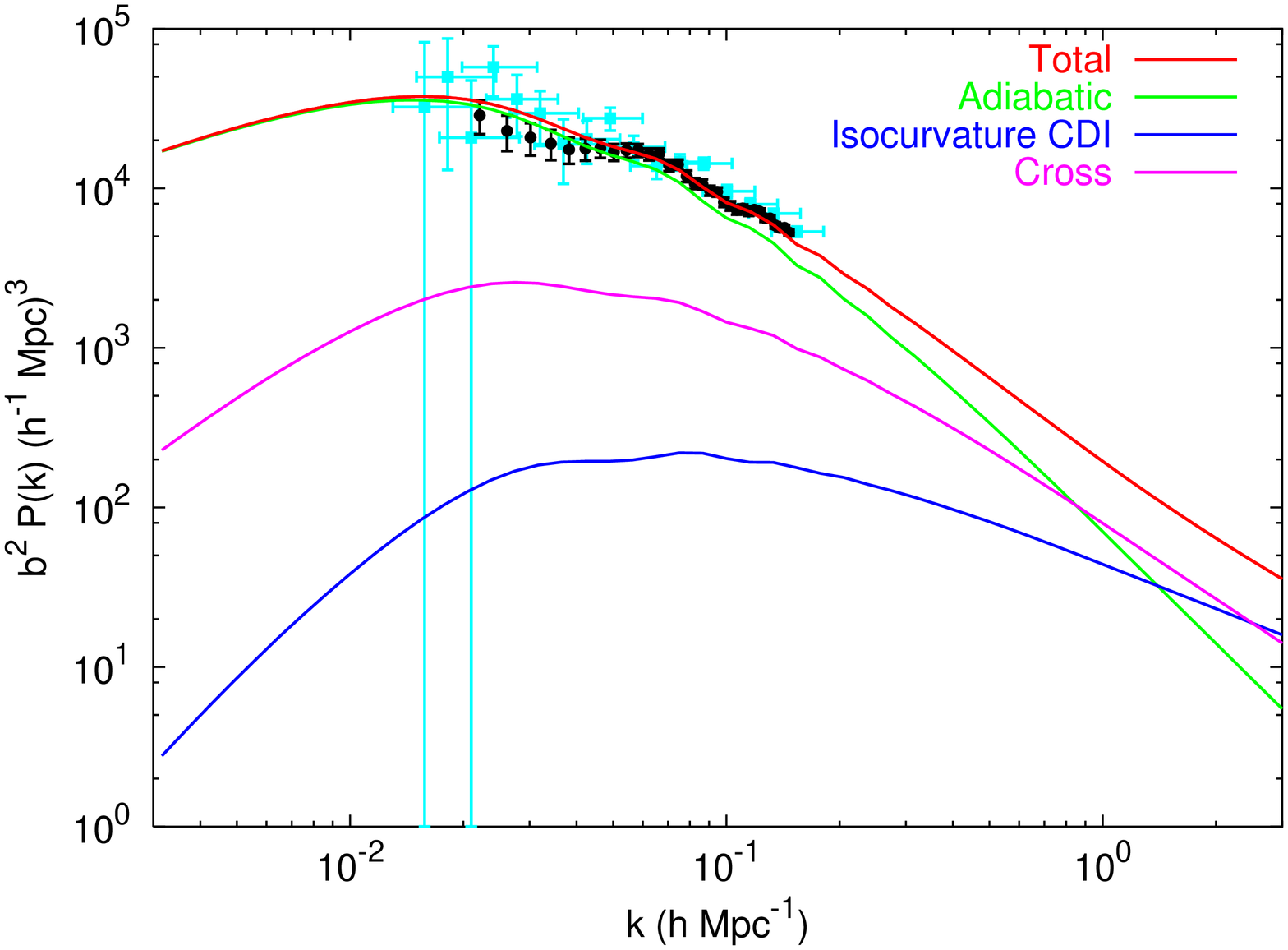}~
\includegraphics[angle=-90,width=8cm]{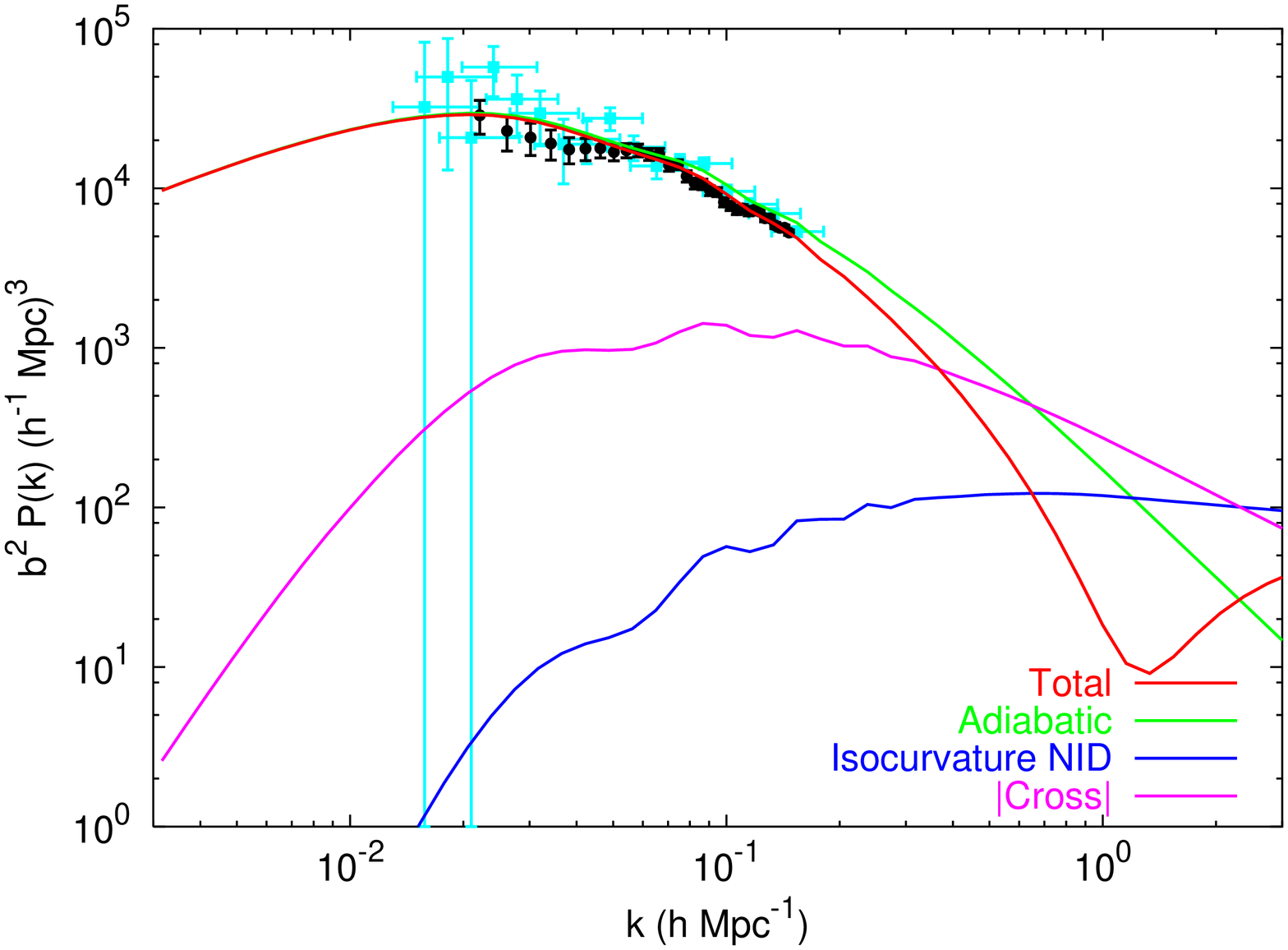}
\end{center}
\caption{\label{fig:spectra} 
Temperature, E-polarization and matter
power spectra for two particular CDI and NID models. In order to get a
better understanding of our bounds, we chose here two models with
large values of $\alpha$, still allowed at the 2-$\sigma$ level:
respectively, $\alpha=0.53$ and $\alpha=0.41$.  Other parameter values
are for the CDI (resp. NID) model: $\oB=0.0217$ (0.0196),
$\ocdm=0.112$ (0.131), $\theta=1.06$ (1.01), $\tau=0.068$ (0.131),
$w=-0.88$ (-1.44), $\nad=0.96$ (1.02), $\niso=2.93$ (2.95),
$\ncor=0.05$ (0.03), $\ln[10^{10}{\cal R}_{\rm rad}]=3.73$ (3.98),
$\beta=-0.62$ (0.88).  The $\chi^2$ of the two models is respectively
1675 and 1674.  From top to bottom, we show the $\cltt$, $\clte$ and
$P(k)$ power spectra, as well as the contribution of each component:
adiabatic, isocurvature, cross-correlated, total. The CDI isocurvature
and cross-correlated components have been rescaled by a
factor indicated in each figure. We also show the data points that we
use throughout the analysis, from WMAP (black), ACBAR (grey), CBI (blue), 
VSA (yellow), 2dF (black) and SDSS (blue).
In the case of the matter power spectrum, one should not trust a
``$\chi^2$-by-eye'' comparison with the data: first, because the
spectrum has to be convolved with the experimental window function
before the comparison (this changes its slope significantly); 
second, because we show here the
data points before rescaling by the two bias factors, which are left
arbitrary for each experiment.
}
\end{figure*}

On Figure~\ref{fig:spectra}, we plot the CMB and LSS power spectra
for two particular CDI and NID models. In order to get a better understanding
of our bounds, we chose models with large values of $\alpha$, still
allowed at
the 2-$\sigma$ level: respectively, $\alpha=0.53$ and $\alpha=0.41$. 
The detailed values of other cosmological parameters
for these models are given in the figure caption. In the CDI example, 
one can see that the non-adiabatic contributions to all spectra remain
tiny, excepted for the matter power spectrum on scales $k>0.2 h\,$Mpc$^{-1}$,
due to the large isocurvature tilt $\niso=2.93$ of the model. 
This is an indication that our $\alpha$ bound in the CDI case depends
very much on constraints on the small-scale matter power spectrum, while
future improvements in the determination of CMB spectra would not reduce
it dramatically. Note that a precise experimental
determination of the amplitude parameter $\sigma_8$ 
(which is mainly sensitive to scales around $k =
0.2\,h$~Mpc$^{-1}$)
would probably
not change things either,
since our CDI models have roughly the same $\sigma_8$ values as 
purely adiabatic well-fitting models (see Figure~\ref{1Dplots_bis});
on the other hand, any constraint on smaller wavelengths
could improve the bounds.
This means in particular that
our choice not to include the Lyman-$\alpha$ data
plays a crucial role in our results.

The same conclusions apply to the NID model of Fig.~\ref{fig:spectra}.
In addition, in the NID case, we see that the non-adiabatic contribution
is significant also for small-scale ($l>200$) temperature and
polarization spectra. Indeed, it is well-known that the NID isocurvature
and cross-correlated modes can mimick the adiabatic CMB spectra to a
better extent than CDI modes (essentially because the amplitude of the
secondary peaks is not strongly supressed). So, in the NID case, future
improvement in the CMB data should help to improve the bounds on the
isocurvature fraction.

Our bounds are difficult to compare with those of
Ref.~\cite{Moodley:2004ws}, because of our different parameter space,
observational data set and conventions of normalization. In the AD+CDI
case, our analysis is closer to the one of the WMAP team \cite{WMAP},
although we have more free cosmological parameters ($n_{\rm cor}$,
$w$), more data (SDSS, CBI, VSA) and less constraints on the
matter--to--light bias. The 95\% limit ${\cal B}<0.33$ obtained by
WMAP would correspond to $\alpha<0.1$ in our notations, which is
significantly smaller than our results. Also, The WMAP $1\sigma$
bounds on $n_{\rm iso}$ are $1.26\pm0.5$, while we find that the
likelihood peaks at our maximum allowed value $n_{\rm iso}=3$.  The
most likely explanation is that the use of the
Lyman-$\alpha$ data in the WMAP analysis eliminates all our well-fitting
models with $n_{\rm
iso} > 2$ and large $\alpha$ values. 
Similar conclusions apply to our previous results \cite{Crotty:2003aa}, in
which we did not use any Lyman-$\alpha$ information, but
adopted a flat prior $0.6 < n_{\rm iso} < 1.5$ (this was the
interval in which our grid of models was computed). Then,
most of the well-fitting models of \cite{Crotty:2003aa} had slighly
negative values of $\nad-\niso$. Therefore, translating our previous
results in terms of a pivot scale $k_0=0.05\,$Mpc$^{-1}$ would lead to
a small decrease in the $\alpha$ bounds, making them comparable
with the WMAP bound in the CDI case, and smaller
than the conservative bounds of this work.

\section{Double Inflation}

In this section we will use the previous bounds to constrain the
parameters of a concrete model of inflation called double inflation
\cite{double,Langlois:1999dw}. This is an inflationary model with two
massive fields, $\phi_1$ and $\phi_2$, of different masses, with ratio
$R=m_1/m_2$. No coupling (except gravitational) between fields is
assumed. The equations of motion and Friedmann equation can be written
as
\begin{eqnarray}
&&\ddot\phi_1 + 3H\,\dot\phi_1 + m_1^2 \phi_1 = 0\,,\\[2mm]
&&\ddot\phi_2 + 3H\,\dot\phi_2 + m_2^2 \phi_2 = 0\,,\\[2mm]
&&H^2 = {\kappa^2\over6}\Big(\dot\phi_1^2 + \dot\phi_2^2 + m_1^2 \phi_1^2 
+ m_2^2 \phi_2^2\Big)\,,
\end{eqnarray}
where $\kappa^2\equiv8\pi G$. In the slow-roll approximation,
$\dot\phi_i^2 \ll m_i^2 \phi_i^2\,; \ddot\phi_i \ll H\dot\phi_i$, we
have the solutions~\cite{double}
$$\phi_1 = \left({4s\over\kappa^2}\right)^{1/2}\,\sin\theta\,;\hspace{5mm}
\phi_2 = \left({4s\over\kappa^2}\right)^{1/2}\,\cos\theta\,,$$
with $s=-\ln(a/a_{\rm end})$ the number of $e$-folds to the end
of inflation. 
For simplicity, we restrict ourselves to the case where $\phi_1$ and
$\phi_2$ remain positive during inflation.
Substituting into the rate of expansion,
\begin{equation}
H^2(s) \simeq {2\over3} m_2^2\,s\,\Big(1 + (R^2-1)\sin^2\theta\Big)\,,
\end{equation}
and, using the equations of motion, one can integrate out
\begin{equation}\label{thetas}
{d\theta\over ds}(s) = {\tan\theta\over2s}\,{R^2-1\over
1+ R^2\,\tan^2\theta}\,.
\end{equation}
to obtain
$$s = s_0\,{(\sin\theta)^{2\over R^2-1}\over
(\cos\theta)^{2R^2\over R^2-1}}\,.$$
As inflation proceeds, $s$ decreases and $\theta$ also decreases.
We will call $s_H=60$ to the number of $e$-folds before the end of
inflation when the scale corresponding to our Hubble radius today
exited during inflation.

\subsection{Linear perturbations}

The great advantage of double inflation is that it is possible to find
explicit formulas for the perturbations on superhorizon scales.  The
growing mode solutions for the scalar fields and the scalar metric
perturbation in the longitudinal gauge, and in the slow-roll
approximation, are given by~\cite{Polarski:1994rz}
\begin{eqnarray}
&&\Phi = -C_1(k){\dot H\over H^2} + C_3(k)
{2(R^2-1)R^2\phi_1^2\phi_2^2\over
3(R^2\phi_1^2 + \phi_2^2)^2}\,,\hspace{5mm}\\[2mm]
&&{\delta\phi_1\over\dot\phi_1} = {C_1(k)\over H} + C_3(k)
{2H\phi_2^2\over R^2\phi_1^2 + \phi_2^2}\,,\\[2mm]
&&{\delta\phi_2\over\dot\phi_2} = {C_1(k)\over H} - C_3(k)
{2HR^2\phi_1^2\over R^2\phi_1^2 + \phi_2^2}\,.
\end{eqnarray}
Since $\phi_1$ and $\phi_2$ are independent uncoupled scalar fields
and essentially massless during inflation, we can use the general
formalism of  section II, and write, in the slow-roll approximation,
\begin{eqnarray}
&&C_1(k) = -{\kappa^2\over2}{H_k\over\sqrt{2k^3}}\Big(
\phi_1\,e_1({\bf k}) + \phi_2\,e_2({\bf k})\Big)\,,\hspace{5mm}\\[2mm]
&&C_3(k) = - {3H_k\over2\sqrt{2k^3}}\left({e_1({\bf k})\over m
_1^2\phi_1} - {e_2({\bf k})\over m_2^2\phi_2}\right)\,,
\end{eqnarray}
where $H_k$ is the rate of expansion when the perturbation of 
wavenumber 
\begin{equation}\label{ks}
k \simeq k_H\,e^{s_H - s_k}\,,
\end{equation}
left the horizon during inflation, where the scale of our present
horizon is $k_H^{-1} = 3000\,h^{-1}$ Mpc.

We will now assume that the light scalar field decays at the end of
inflation into the ordinary particles, giving rise to photons,
neutrinos, electrons and baryons, while the cold dark matter (CDM)
arises from the decay of the heavy field. In principle, part of the CDM
could also be produced by the light field or the heavy field could also
decay into ordinary particles, but we will ignore these possibilities
here. Then, the perturbations in the comoving gauge take the form
$${\delta^{(c)}n_\gamma\over n_\gamma} = 
{\delta^{(c)}n_\nu\over n_\nu} = 
{\delta^{(c)}n_B\over n_B}\,,$$
and there is only one isocurvature mode, the CDI mode,
$$\calS \equiv \delta^{(c)}\ln{n_{\rm cdm}\over n_\gamma} =
\delta^{(c)}_{\rm cdm} - {3\over4}\delta^{(c)}_\gamma\,,$$ all of
which are gauge invariant quantities.  During the radiation era, the
initial conditions of all these modes are described in terms of only
two $k$-dependent quantites, $\Phi_k$ and $S_k$. The pure adiabatic 
initial conditions are given by the gravitational potential during the
radiation era,
$$\Phi_{\rm rad}(k) = {2\over3}\,\Rrad(k) = {2\over3}\,C_1(k)\,,$$ with
$C_1(k)$ the amplitude of the growing adiabatic mode during
inflation. On the other hand, the isocurvature initial conditions in the
radiation era arise from the perturbations in the heavy field $\phi_1$
at the end of inflation. In the long wavelength limit, the perturbations
of this field during reheating follow closely the field itself, so that
its energy density perturbations satisfy, in the comoving gauge,
$${\delta\rho^{(c)}\over \rho} = 2{\delta\phi_1\over\phi_1} =
-{4\over3}C_3(k)\,m_1^2$$
which is constant during inflation, across reheating and into the
radiation era. The entropy perturbation is dominated by the CDM
density perturbation during the radiation era, $\Srad \simeq
\delta^{(c)}_{\rm cdm}$. Using the values of 
$C_1(k)$ and $C_3(k)$ during inflation, we can finally write
\begin{eqnarray}
&&\Rrad(k) = -{\kappa H_k\over\sqrt{2k^3}}
\sqrt{s_k}\Big(\sin\theta_k\,e_1({\bf k}) + 
\cos\theta_k\,e_2({\bf k})\Big)\,,
\hspace{5mm}\\[2mm]
&&\Srad(k) = {\kappa H_k\over\sqrt{2k^3}}{1\over\sqrt{s_k}}
\left({e_1({\bf k})\over \sin\theta_k} - {R^2\,e_2({\bf k})\over 
\cos\theta_k}\right)\,.
\end{eqnarray}
Note that the cross-correlation amplitude $\cos\Delta(k)$ is almost
independent of $k$, due to the cancellation of the factor $s_k$
between $\Rrad(k)$ and $\Srad(k)$, and to the
fact that $\theta_k$ is a mild function of $k$, see Eq.~(\ref{thetas}).
In this case, the tilt $\ncor$ vanishes. 
The adiabatic and isocurvature tilts can be computed explicitly 
using~(\ref{ks})
\begin{eqnarray}
&&\calPR(k) = {k^3\over2\pi^2}\langle|\calR_k|^2\rangle\,, 
\hspace{5mm} 
\calPS(k) = {k^3\over2\pi^2}\langle|\calS_k|^2\rangle\,, \ \\[2mm]
&&\nad = 1 - {\partial\ln\calPR(k)\over\partial\ln s}\,, 
\hspace{2mm} 
\niso = 1 - {\partial\ln\calPS(k)\over\partial\ln s}\,,
\hspace{5mm}
\end{eqnarray}
as functions of the angle $\theta$ and the mass ratio $R$, 
\begin{eqnarray}\label{nad}
\nad \!&=\!& 1 - {2\over s} + {(R^2-1)\tan^2\theta\over
2s\,(1+R^2\tan^2\theta)^2}\,,\\[2mm]
\niso \!&=\!& 1 - {(R^2-1)(R^6\tan^4\theta-1)\over
s(1+(R^2-1)\sin^2\theta)}\times\\[2mm]\label{niso}
&&\hspace{1cm}{1\over(1+(R^2+R^4)\tan^2\theta+R^6\tan^4\theta)}
\end{eqnarray}

\begin{figure}[ht]
\hspace*{-2mm}
\includegraphics[width=6cm,angle=-90]{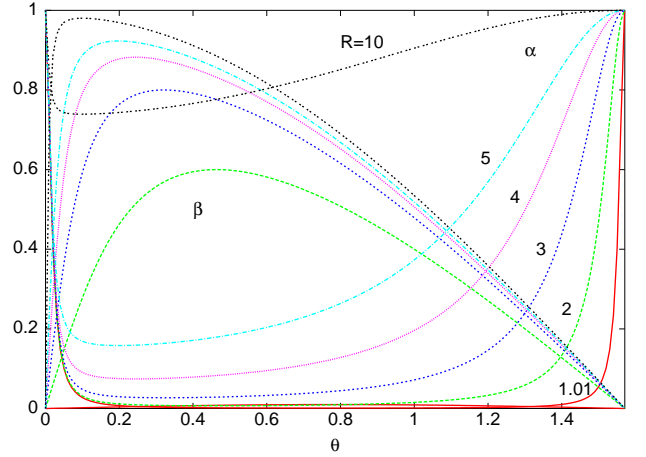}
\caption{The values of parameters $\alpha$ and $\beta$ as 
a function of the angle $\theta$, for different values of the ratio 
$R=m_1/m_2$ in double inflation. 
\label{fig:alphabeta}}
\end{figure}

>From $\calPR$ and $\calPS$, we obtain the parameters
\begin{eqnarray}\label{alpha}
&&\alpha = {R^4\tan^2\theta +1\over s_k^2\sin^2\theta + 
R^4\tan^2\theta +1}\,,\hspace{5mm}\\[2mm]\label{beta}
&&\beta = {(R^2-1)\sin\theta\over\sqrt{R^4\tan^2\theta +1}}\,.
\end{eqnarray}
We have plotted these parameters as a function of the angle $\theta$
in Fig.~\ref{fig:alphabeta}, for $s_k \simeq 60$. 

\noindent
These parameters 
have maximum and minimum values given by
\begin{eqnarray}
&&\alpha_{\rm min} = {(R^4-1)^3\over(R^4-1)^3 + s_k^2\Big[R^8 - 2R^2
(R^4-1) - 1\Big]}\,,\hspace{5mm}\\[2mm]
&&\beta_{\rm max} = {R^2-1\over R^2 +1}\,,
\end{eqnarray}
as can be seen in Fig.~\ref{fig:alphabeta}.

\subsection{Bounds on double inflation}

In this subsection we will impose the general bounds found in section
III to the double inflation model~\cite{double}.  Note that in this
model it is assumed that the heavy field decays into cold dark matter,
and therefore we only have one isocurvature component, CDI. The bounds
from CDI will be used to constrain this particular model. We will
leave for the future a detailed analysis of other two-field models of
inflation.

In order to derive specific constraints, it will be useful to take
into account the following relation between $\alpha$ and $\beta$,
see Eqs.~(\ref{alpha}) and (\ref{beta})
\begin{equation}
2\beta\sqrt{\alpha(1-\alpha)} = {R^2-1\over s_k/2}\,(1-\alpha)\,,
\end{equation}
which corresponds to a straight line in the contour plot of
Fig.~3. This way, one can evaluate the likelihood at which a given
value of $R$ is ruled out. Unfortunately, from the
contours in Fig.~3, one cannot restrict much the range of $R$, except
to exclude $R>5$ at 2$\sigma$.

Even if the model passes this constraint for a given $R$, it is possible
that the prediction on $\nad$ (\ref{nad}) and $\niso$ (\ref{niso}) do
not agree with the bounds on these parameters.  In our case, the bounds
are so loose that any tilt is allowed.  Perhaps in the future, with
better observational constraints, we may use the information on the
tilts to further rule out double inflation models.

\subsection{Massive complex field}

Another model worth exploring is the particular one in which
the two fields have equal masses, corresponding to a massive complex
field $\Phi = {1\over\sqrt2}(\phi_1 + i\,\phi_2)$. If we rewrite
$\Phi={1\over\sqrt2}\sigma\,\exp(i\varphi)$, with modulus $\sigma$ and 
phase $\varphi$,
the lagrangian can be written as
$${\cal L} = \half(\partial_\mu\sigma)^2 + \half\sigma^2(\partial_\mu
\varphi)^2 - \half m^2\sigma^2$$
Note that there is no potential for the phase, so it will be free to
fluctuate, which will induce a large isocurvature component, as we
will see, and which can be used to rule this model out.

In this case, the curvature and entropy perturbations are
\begin{eqnarray}
&&\Rrad(k) = -{\kappa H_k\over\sqrt{2k^3}}\sqrt{s_k}\,e_\sigma({\bf k})\,,
\hspace{5mm}\\[2mm]
&&\Srad(k) = -{\kappa H_k\over\sqrt{2k^3}}{1\over\sqrt{s_k}}\,
e_s({\bf k})\,,
\end{eqnarray}
with $e_\sigma$ and $e_s$ orthonormal. Therefore,
\begin{equation}
\alpha = {1\over s_k^2 + 1}\,,
\hspace{1cm}\beta=0\,.
\end{equation}

The curvature perturbation has a tilt $\nad = 1-2/s = 0.97$, but the
isocurvature perturbation has no tilt, $\niso = 1$, and the two modes
are uncorrelated, $\beta=0$. 
For the moment, this model is not ruled out.

\section{Conclusions}

Using the recent measurements of temperature and polarization
anisotropies in the CMB by WMAP, together with recent data from VSA,
CBI, and ACBAR; the matter power spectra from the 2dF galaxy redshift
survey and the Sloan Digital Sky Survey, as well as the recent
supernovae data from the SN Search Team, one can obtain stringent
bounds on the various possible isocurvature components in the
primordial spectrum of density and velocity fluctuations. We have
considered correlated adiabatic and isocurvature modes, and find no
significant improvement in the likelihood of a cosmological model by
the inclusion of an isocurvature component, see Table~1 and
Fig.~3. So, the pure adiabatic scenario remains the most
economic and attractive scenario. 

In contrast with the WMAP analysis,
we decided not to include any data from Lyman-$\alpha$ forests, since
constraints on the linear power spectrum coming from these experiments
are derived under the assumption of a plain adiabatic $\Lambda$CDM
scenario. We did not include either strong priors on the
isocurvature spectral index, unlike
in our previous work, and allowed this parameter to vary up to $\niso=3$.
This conservative approach leads to a preference for
models with a very blue isocurvature primordial spectrum, and to upper
bounds on the
isocurvature fraction significantly larger than in other recent 
analyses: on a pivot scale
$k=0.05\,h\,$Mpc$^{-1}$ the amplitude of the correlated isocurvature
component can be as large as about 60\% for the cold dark matter
mode, 40\% for the neutrino density mode, and 30\% for the neutrino
velocity mode, at 2$\sigma$. This leaves quite a lot of freedom, for instance,
for double inflation models with two uncoupled massive fields. Assuming that
one of these fields decay into Cold Dark Matter, our results simply
imply that the mass of the heavy field cannot exceed five times that of the
light field at the 2$\sigma$ confidence level.
It is
expected~\cite{Trotta:2004sm} that in the near future, with better
data from Planck~\cite{Planck} and other CMB experiments, we will be
able to reduce further a possible isocurvature fraction, or perhaps
even discover it. The present results also suggest that constraining
the linear matter power spectrum on scales which are mildly non-linear
today will also
be crucial in this respect.

\section*{Note added}
After completing this work, we came upon the paper of Parkinson et 
al.~\cite{Parkinson}, where a similar analysis was done for a class
of hybrid inflation models. 

\section*{Acknowledgments}
We thank Andrew Liddle for stimulating discussions.
M.B. thanks the group at Sussex University for their warm hospitality
and acknowledges support by a Marie Curie Training Site Fellowship.
We also acknowledge the use of the COSMOS cluster for our computations
with the CosmoMC code, and thank the sponsors of this UK-CCC facility,
supported by HFCE and PPARC. This research was conducted in
cooperation with GSI/Intel utilising the Altix 3700 supercomputer.
This work was supported in part by a CICYT project FPA2003-0435, and
by a Spanish-French Collaborative Grant bewteen CICYT and IN2P3.

\end{document}